# Technologies to Capture CO₂ directly from Ambient Air

Gahyun Annie Lee[1,3,†], Xiaoyang Shi,[1,2,3†] Ah-Hyung Alissa Park[1,2,3,*]

[1] Department of Chemical Engineering, Columbia University, New York, NY 10027, USA

[2] Department of Earth and Environmental Engineering, Columbia University, New York, NY 10027, USA

[3] Lenfest Center for Sustainable Energy, The Earth Institute, Columbia University, New York, NY 10027, USA

[†] These authors contributed equally

*Corresponding authors: ap2622@columbia.edu

## Contents

## Abstract:

Building a carbon-neutral world needs to remove the excess $CO_2$ that has already been dumped into the atmosphere. The sea, soil, vegetation, and rocks on Earth all naturally uptake $CO_2$ from the atmosphere. Human beings can accelerate these processes in specific ways. The review summarizes the present Direct Air Capture (DAC) technology that contribute to Negative Emissions. Research currently being done has suggested future perspectives and directions of various methods for Negative Emission. New generations of technologies have emerged as a result of recent advancements in surface chemistry, material synthesis, and engineering design. These technologies may influence the large-scale deployment of existing $CO_2$ capture technologies in the future.

# 1. Introduction:

The entire structure of creation depends heavily on carbon. It can be found as dissolved forms in the atmosphere, dead organic debris, fossil fuels, rocks, tissues of plants and animals, and the ocean. The concentration of $CO_2$ in the atmosphere has increased by more than 100 ppm in the past 200 years. The disposed carbon dioxide ($CO_2$) from fossil fuels in the atmosphere is the main factor contributing to climate change[1]. Aggravated climate change increases the probability of severe ecological impacts[2-4]. Current total $CO_2$ emissions from fossil fuels approach 40 Gt/year[5]. Intergovernmental Panel on Climate Change (IPCC) reported $CO_2$ emissions will rise from existing level of 49 $GtCO_2eq/yr$ to between 85 and 136 $GtCO_2eq/yr$ by 2050 without intervention[6]. The rising $CO_2$ concentration might raise the earth's average temperature from 3.8 to 6.0 °C by 2100 compared to the pre-industrial level[6]. It will take some time for the energy infrastructure to transition from using fossil fuels to using renewable energy sources. Utilizing localized $CO_2$ capture and storage (CCS) systems, power stations and industrial producers might minimize $CO_2$ emissions while burning fossil fuels[7-10]. However, lowering $CO_2$ emissions from solitary point sources will not be enough to reach the objective of keeping atmospheric $CO_2$ at 450 ppm and reaching carbon neutrality before 2050[11,12]. This will necessitate the development of technology to directly capture and permanently remove $CO_2$ from ambient air[13-19]. Capturing $CO_2$ from the atmosphere, referred as Direct Air Capture (DAC), has the potential to lessen or perhaps stop global warming[20,21].

Since the early report on evaluating the feasibility of Direct Air Capture system by American Physical Society in 2011[22], it is asserted that DAC is already capable of capturing 1 percent of annual world $CO_2$ emissions[23]. The cost estimation of DAC has been reduced from the ranges of 610 to 1100 $/ton $CO_2$ [22,24] to 94 to 232 $/ton $CO_2$[25]. Considering the thermodynamic system efficiencies and environmental impact of the technology, life cycle assessment (LCA) is an efficient multi-step process for evaluating the feasibility of large-scale DAC deployment[26-40]. Pielke[41] calculated the cost of DAC as a percentage of global GDP to 2100 for $CO_2$ stabilization at 450 and 550 ppm. The cost of DAC is in the range between 0.3% and 3% of global GDP. The generated LCA results vary depending on the capture methods used, the energy sources used for system regeneration, and other factors like as climate and environmental consequences, air pollution, and societal cost. Overall, the LCA analysis demonstrates the feasibility of using DAC as a climate-mitigation technique.

Increasing number of policies passed or enacted in recent years supporting DAC industries in the U.S [42]. The U.S. Department of Energy (DOE) launched Carbon Negative Earthshots in November 2021 to accelerate the goal of net-zero emissions by 2050, hoping to deploy Carbon Dioxide Removal (CDR) at a gigaton scale. The Carbon Negative Earthshots is one of the first U.S government's major efforts in CDR to tackle the imminent climate crisis global society faces. The new targets aim to decrease the total cost of CDR to lower than $100 per ton by 2050. DOE will spend approximately $10 billion on new direct carbon management funding through the Bipartisan Infrastructure Law. In addition, most recently passed the Inflation Reduction Act (IRA) in August 2022 is expected to substantially increase the tax credits toward the operations that qualify under specific high standards for wage and workforce conditions: for example, $180/ton for sequestered $CO_2$ and $130/ton for capture-and-reuse systems[43,44]. Accordingly, strengthened climate goals with mandates and financial investment incentives are delivering unprecedented momentum for DAC to be viable[43]. With federal support, we believe CCUS technologies will play an essential role in meeting net zero targets, including point source capture, which tackles emissions from heavy industry and direct capture to remove carbon from the atmosphere[45-47]

## 2. Technologies of Direct Air Capture of $CO_2$

Direct Air Capture (DAC) is a technology-based solution for carbon capture. Lackner first introduced to capture $CO_2$ from the air, or DAC, to fight the climate change in 1999[48]. Researchers argued Direct Air Capture of $CO_2$ is a necessary and practical way to reduce $CO_2$ concentration in the atmosphere [25,26,41,49-76]. IPCC claimed negative emission technologies, such as DAC, is necessary to solve climate change issue[77]. The economic feasibility of DAC has also been established[23,25,78-83].

A few companies pioneered this effort. Carbon Engineering[25] uses a hydroxide solution to capture $CO_2$, and then heat sorbent to high temperatures to release $CO_2$ for storage. The business model relies on selling captured $CO_2$ to make new products. Climeworks[84] based out of Switzerland uses amine sorbents in small modular reactors that cost more than Carbon Engineering, but the potential savings down the road could be higher. The idea is that the modular design could make the process cheaper to produce when scaled up. It also requires lower temperatures during the regeneration process by using waste heat. A similar model is being followed by another company, Global Thermostat, which partners up with industrial plants and uses their waste heat to capture $CO_2$ from the air and offsets the emissions from that plant. Verdox uses an electrochemical approach[85] to capture $CO_2$ from industrial sources and the air. Carbon Collect employs mechanical trees to capture and release $CO_2$ by controlling moisture[59,86]. Captured $CO_2$ by DAC technology can be utilized for various purposes in industry, such as enhanced oil recovery, beverage manufacturing, concrete curing, food transportation, *etc*.

The most significant portion of DAC technology is the advance of energy-saving materials to capture and release $CO_2$. Currently, there are seven types of DAC technologies, including physical sorption[87-94], strong bases sorption[49,51,87,95-101], amine-functional materials[9,102-110], amino acid solution and guanidine compound[111-119], moisture-swing sorption[14,59,86,120-133], and electrochemical sorption[85]. The six typical technologies for DAC are shown in Figure 1.

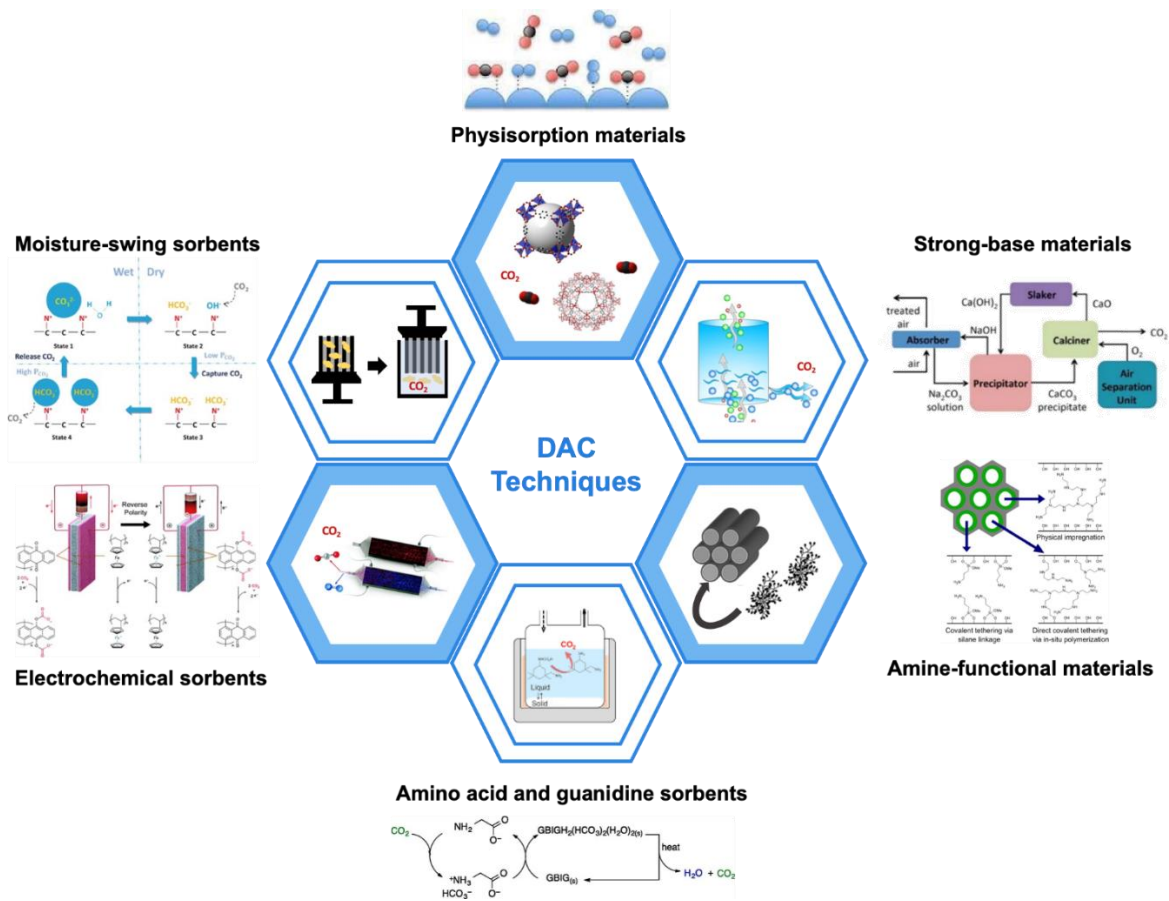

Figure 1. Six standard technologies for Direct Air Capture of $CO_2$ (DAC).

## 2.1 Physisorption Materials:

Van der Waals or ion quadrupole forces determine the performance of $CO_2$ capture by physisorption[88]. Physisorption requires a considerably low energy of regeneration less than 60 kJ/mol compared to chemisorption above 70 kJ/mol[134], shown in Figure 2. The regeneration process Physisorption is convenient and sustainable. However, the competitive physisorption by $H_2O$ in the air may jeopardize the capacity of sorbents. A material with a high surface-to-volume ratio is required for physisorption of $CO_2$[89-91]. Potential candidates include metal-organic frameworks (MOFs), zeolites, alumina, nanostructured graphite, and so on[87,135,136].

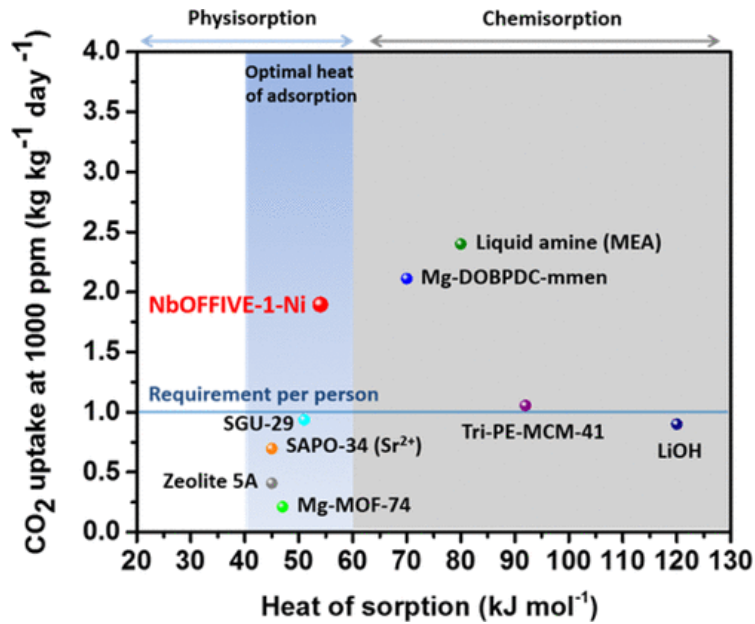

Figure 2. Heat of adsorption-$CO_2$ uptake (at 1000 ppm for 1 day).[134]

Madden et al. [135,136] studied the $CO_2$ sorption capacities of different porous materials, including zeolite, amine-impregnated SBA-15, and MOFs, as shown in Figure 3. The study claims that water in the atmosphere can significantly reduce their $CO_2$ capture performance, due to the competitive sorption and the stability issue of MOFs in a moisture condition.

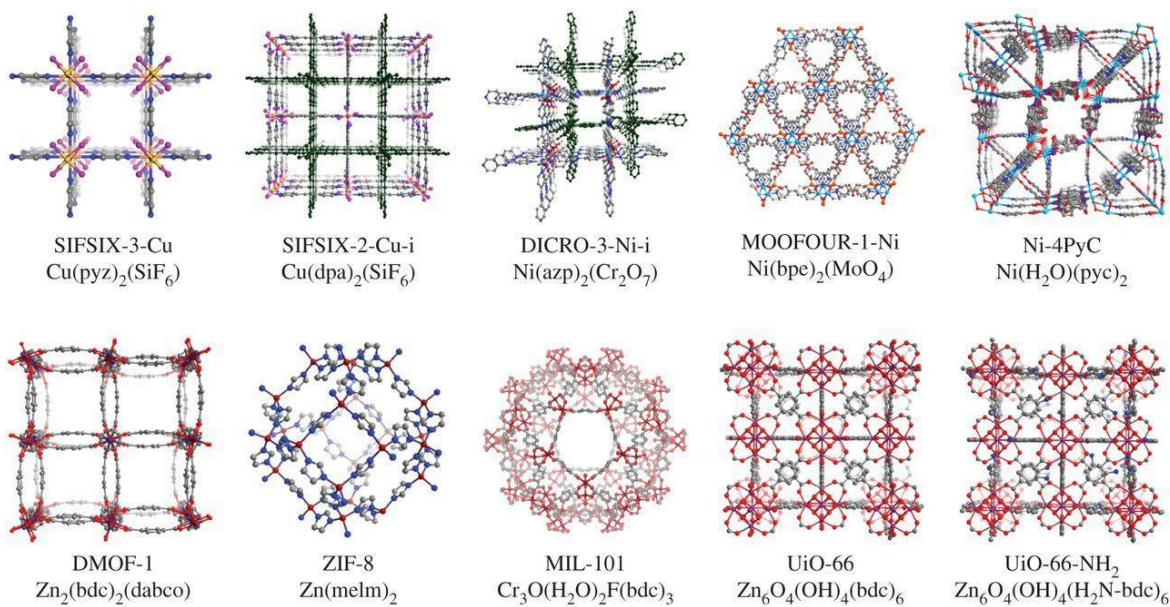

Figure 3. Molecular structures of MOFs used as $CO_2$ sorbents by physisorption[136]

Mukherjee *et al.*[137] summarized the $CO_2$ capacity and kinetics of six types of MOFs under low concentration of $CO_2$ with dry air or wet air (74% relative humidity). NbOFFIVE-1-Ni and TIFSIX-3-Ni show the highest $CO_2$ capacity of 1.3 mmol/g and 1.2 mmol/g under 500 ppm of $CO_2$ in the dry air condition, respectively. The reason is the adjustment of pore size can significantly affect the performance of $CO_2$ sorption by MOFs materials. Figure 4 shows humidity decreases $CO_2$ capacity and kinetics of the five sorbents. Authors also studied the decrease of $CO_2$ capacity from 2-10% over six consecutive sorption-desorption cycles.

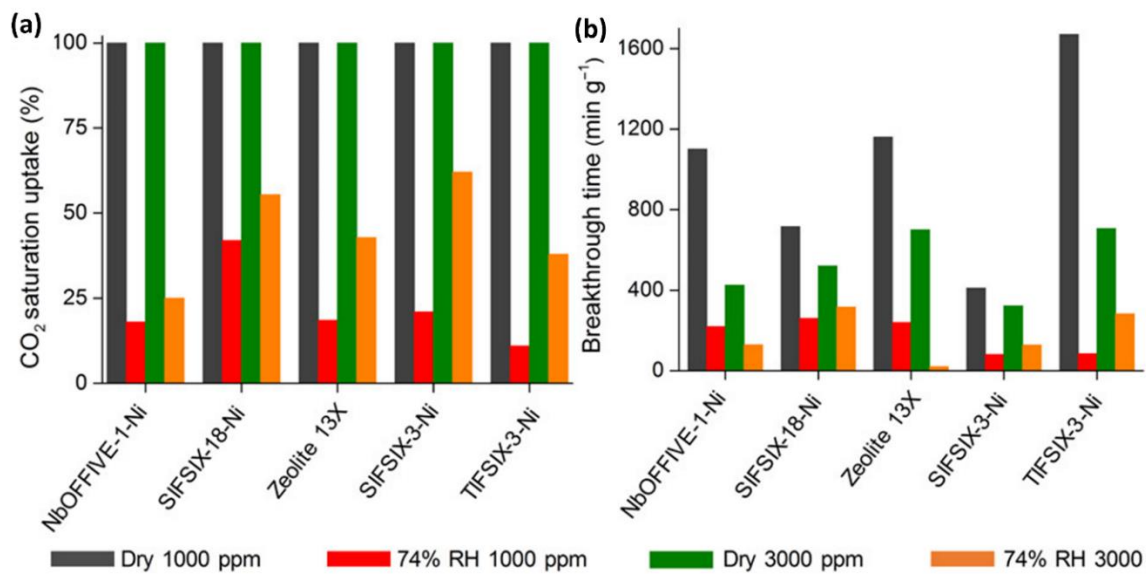

Figure 4. The performance of $CO_2$ sorbent under $CO_2$ concentration of 1000 ppm and 3000 ppm in the dry and wet conditions. (a) $CO_2$ capacity of five sorbents. (b) $CO_2$ retention time.

## 2.2 Strong-base Materials:

$CO_2$ chemisorption is different from physisorption because of the bond break and formation between the $CO_2$ and sorbent materials. $CO_2$ can be easily removed from the air by metal hydroxides or alkaline salts, and then converted to metal carbonates during chemisorption[49,51,87,95-100,138-141]. A calcination procedure can then be employed to regenerate the sorbent and produce a concentrated stream of $CO_2$.[142,143]. However, the calcination process is energy-intensive requiring high temperatures input. Keith *et al*.[25] reported the energy consumption of the process chemistry and thermodynamics used in Carbon Engineering, shown in Figure 5. Overcoming the high energy consumption and heat losses during heating materials are challenging to prevent[144-146], which makes the strong-base materials for $CO_2$ capture from air technically and economically difficult.

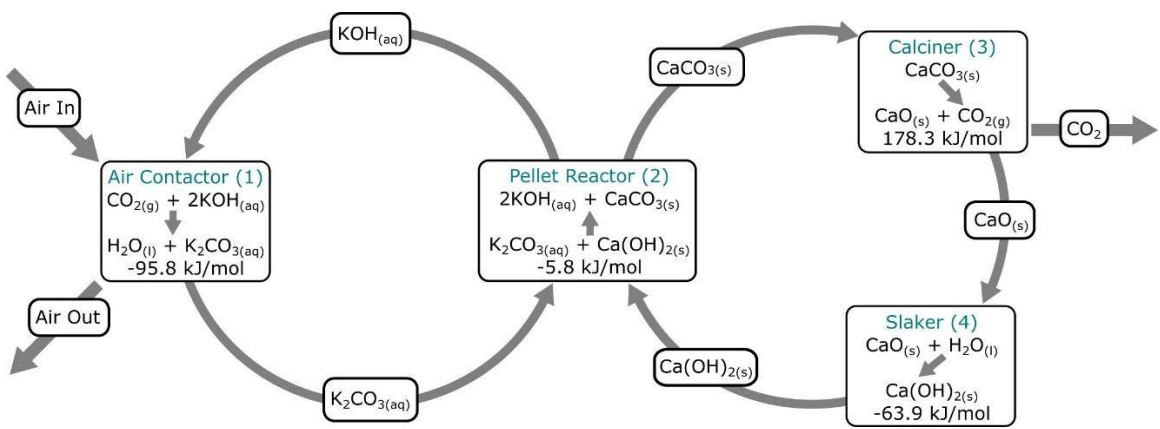

Figure 5. Process chemistry and thermodynamics[25]

## 2.3 Amine-functional Materials:

The most popular way to prepare solid amine-functional sorbents include (1) impregnating amines in porous materials. Polyethylenimine (PEI) was first loaded into a mesoporous molecular sieve as a $CO_2$ sorbent in 2002[147]. In the following decades, researchers impregnated amines into different porous materials, such as alumina, silica, zeolites, nanostructured graphite, black carbon, MOFs, COFs, and so on[147-166] (2) Adding the amine group chemically to the substrate surface[167-172]. Youssef Belmabkhout *et al.*[167] were the first to suggest using triamine-grafted mesoporous silica, TRI-PE-MCM-41, to capture $CO_2$ from the air. According to Mattew E. Potter et al.[173], Because alumina supports are more resistant to moisture and air than silica supports are, they may provide possible stability benefits over porous silica. (3) Amine tethering and ligand modification[167-172,174-177]. The above three methods are shown in Figure 6.

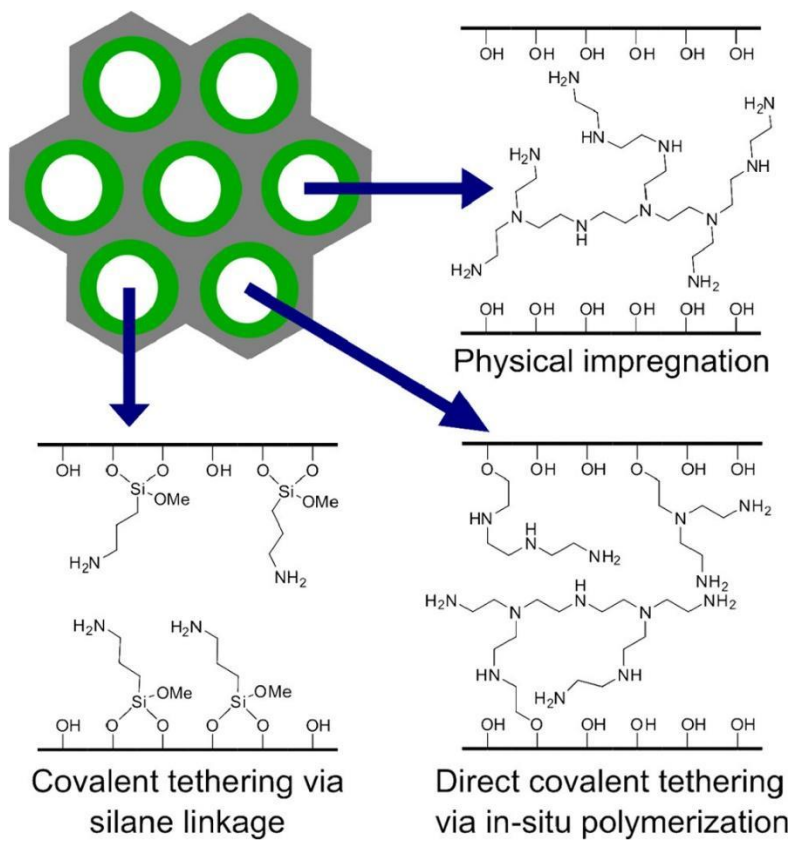

Figure 6. The three main routes for functionalizing porous supports with amine moieties. [13]

Two types of amine-functionalized MIL-101(Cr) were explored by Darunte *et al.*[178] : MIL-101(Cr)-TREN and MIL-101(Cr)-PEI-800. In this study, a small polyamine with high primary amines density, tris (2-amino ethyl) (TREN), was loaded to MIL-101(Cr) via both grafting and impregnation, and a large branched poly(ethylene imine) (PEI-800) was loaded into the MIL-101(Cr) MOF via physical impregnation method. What they found was that the grafted MIL-101(Cr)-TREN only had 0.35 mmol/g MOF and impregnated MIL-101(Cr)-TREN had a significant capacity loss from a cyclic test due to the high volatility of TREN under Direct Air Capture condition coupled with temperature swing adsorption. Unlike a small polyamine TREN, PEI-loaded MIL-101(Cr) improved cyclic stability attributed to the lower volatility. The effect of PEI loading was also investigated in this study. Although the amine efficiencies were enhanced as amine loading increased, 85 wt% of PEI loading was found to be optimal considering the trade-offs between capture capacity and kinetics.

Zhu *et al.* [179] have also functionalized Mg-Al mixed metal oxides with branched PEI via the physical impregnation method. They found that using mixed metal oxides (MMOs) increased the mesopores of the materials formed from the exfoliated nanosheets and the calcination process. The authors observed that the PEI-loaded MMOs slightly increased in $CO_2$ uptake when they underwent desorption above 200 °C, while PEI-loaded SBA-15 had a rapid decrease, which is a commonly faced problem of class 1 adsorbents. This study's optimal loading of PEI was 67 wt% before the PEI covered the MMO nanosheets. The authors argue that the optimized $CO_2$ diffusion channel and increased surface area led to improved kinetics, resulting in a higher $CO_2$ uptake of PEI67/$Mg_{0.55}$Al-O. Moreover, the facilitated electrostatic attraction between the uniformly dispersed impregnated PEI and MMO layers led to the thermal robustness of PEI67/$Mg_{0.55}$Al-O composites as well as the accessibility of the $CO_2$ molecules.

Nanoparticle organic hybrid materials (NOHMs) are a new type of self-suspended nanoparticle system[180,181]. Park's group firstly explored the solvating properties of NOHMs for $CO_2$ capture[182], and synthesized five different NOHMs using silica nanoparticles as cores grafted by amines[183] in 2011, following extensive characterization experiment to understand the properties and thermal stability of the material[184-188]. Rim *et al.*[189] successfully encapsulated optimal loading of 49 wt% the highly viscous liquid-like NOHMs as microdroplets size into highly gas permeable polymer, TEGO Rad 2650. They observed approximately 50 times enhanced capture kinetics compared to the bulk NOHMs under 1 atm $CO_2$. The performance of the optimal encapsulated NOHMs under Direct Air Capture condition was 0.88 mmol $CO_2$/g sorbent in 1 h. However, there still remains a challenge to reach the Department of Energy's target kinetics of 1 mmol $CO_2$/g sorbent in 1 h.

### 2.4 Amino Acid and Guanidine Sorbents:

Custelcean *et al.*[111-119,190,191] describe a method for DAC that relies primarily on amino acid solution and guanidine compound materials. First, aqueous amino acid solutions that are widely accessible are used for $CO_2$ sorption. Following that, guanidine compound is used to react with $CO_2$-loaded solution, and the resulting crystallization of an extremely insoluble carbonate salt regenerates the amino acid sorbent. Finally, relatively mild heating of the carbonate crystals results in effective $CO_2$ release and regeneration of the guanidine compound. The process is shown in Figure 7.

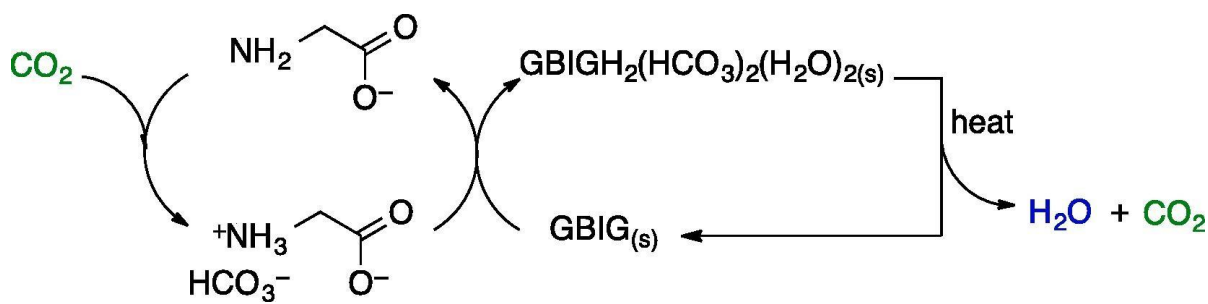

Figure 7. $CO_2$ capture from ambient air using glycine and guanidine[116].

The kinetics of crystallization, particle size distribution, and crystal habit are only a few of the many factors that must be adjusted to optimize the DAC technology.

## 2.5 Moisture-swing $CO_2$ Sorbent:

Lackner[59] advocated employing moisture-swing sorbents to capture $CO_2$ from ambient air with lower energy cost compared with amine method. Water is essential in the $CO_2$ sorption-desorption process[122,129,192]. Moisture-swing $CO_2$ sorbent uptakes $CO_2$ in dry environments and release it in moisture. The sorbent is an anion-exchange resin that has carbonate ions and quaternary ammonium cations permanently attached to polystyrene. Shi *et al.*[86] elucidated the sorption and desorption mechanisms. Other sorbents for the DAC approach were created by impregnating carbon black with an amine-containing polymer[193-195], developed from a biomass material[123], prepared by bamboo cellulose[127].

Armstrong[131] investigated the availability and kinetic uptake of $CO_2$ in sorbents encased in different matrices. A commercially available industrial film containing ion-exchange resin (IER), IER particles embedded in dense electrospun fibers, and IER particles embedded in porous electrospun fibers are compared using a solvothermal polymer additive removal technique to create porosity in the porous fibers. The experimental results indicate that electrospinning polymer/sorbent composites is a promising technology for improving the handleability of sorbent particles and the sorption kinetics, with the IER embedded in porous electrospun fibers demonstrating the highest cycle capacity with an uptake rate of 1.4 mol of $CO_2$ per gram-hour, shown in Figure 8.

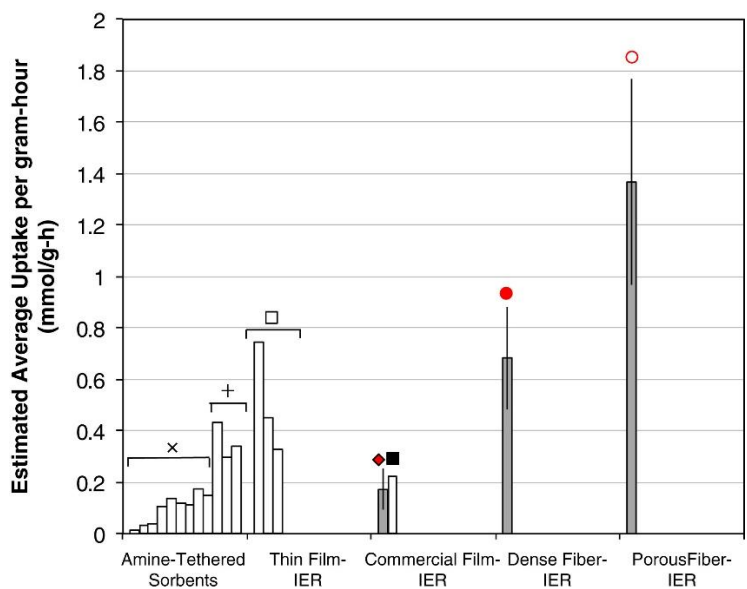

Figure 8. The estimated average amount of carbon dioxide absorbed per gram per hour by sorbents during direct air capture.

Shi *et al*. recently reported on the effects of sorbent parameters on $CO_2$ capture efficiency, which may pave the way for future moisture swing technology-based optimization of various materials for DAC, shown in Figure 9 [86]. Moisture swing performance is weather dependent, and it works best in hot, dry environments.

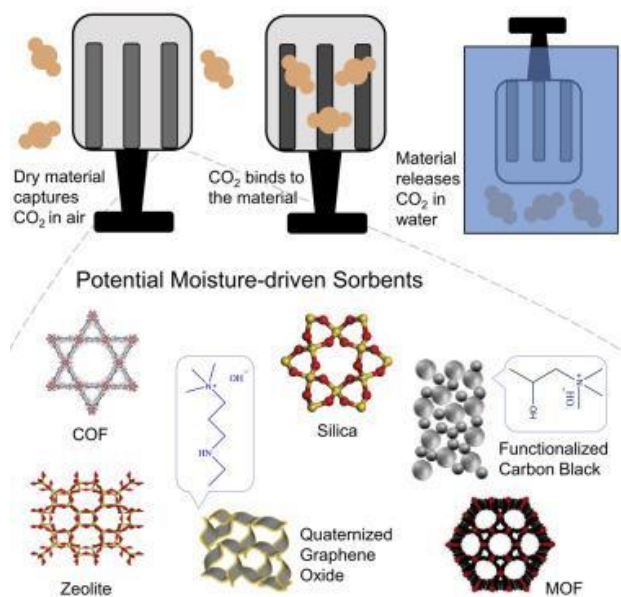

Figure 9. Material candidates for $CO_2$ capture by moisture-swing technology[86].

## 2.6 Electrochemical Sorbent:

The creation of innovative, cost- and energy-efficient solutions is still a top focus for DAC research. Recently, interest in electrochemical $CO_2$ capture has increased due to the fact that it does not require heat energy to release $CO_2$[196-198]. Several electrochemical-swing reactive sorption systems were recently published in Hatton's group[85,199-201]. Voskian *et al*. firstly demonstrated a solid-state electro-swing reactive adsorption system that uses an electrochemical cell to absorb carbon dioxide through the reductive addition of $CO_2$ to quinones[85]. Then, an electrochemically mediated DAC system composed by stackable bipolar cells is described that relies solely on the electrochemical voltage for $CO_2$ capture and release by quinone chemistry[199]. The energy consumption was reported as low as 113 kJ/mol of $CO_2$ captured. Diederichsen et al.[200] further advanced the method of electrochemically mediated carbon capture with high-concentration liquid quinone chemistry in a continuous two-cell flow system. The system is shown in Figure 10.

In a whole bench scale process, the device managed to gather and release $CO_2$ continuously while maintaining high electrochemical stability. The electrochemical process can be further optimized by the electrode and electrolyte chemistry. The stability of quinones during air oxidation and Faradaic efficiency during the capture condition of 400 ppm $CO_2$ can be further improved in future studies. Further research into electrochemically capturing $CO_2$ directly from the air is required.

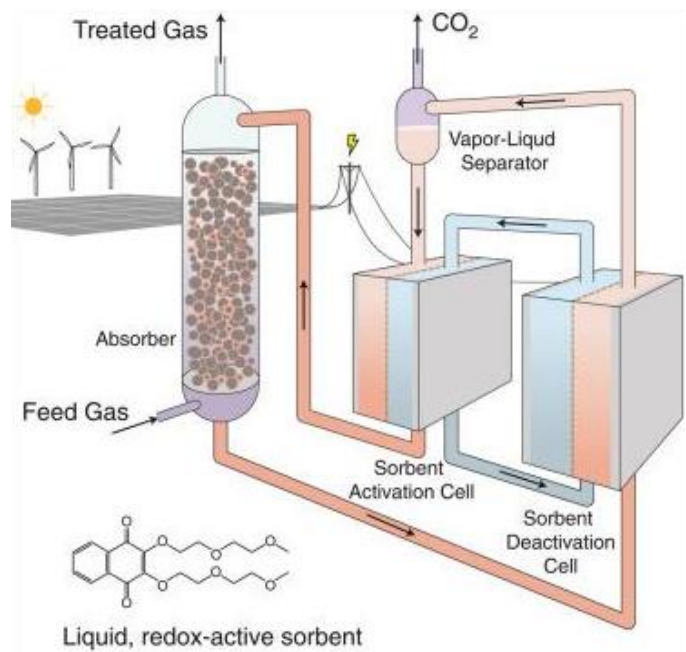

Figure 10. A concept of a continuous two-cell flow system[200]

CO₂ uptake capacities and kinetics of all the types of sorbents are summarized in Figure 11

**(a)**

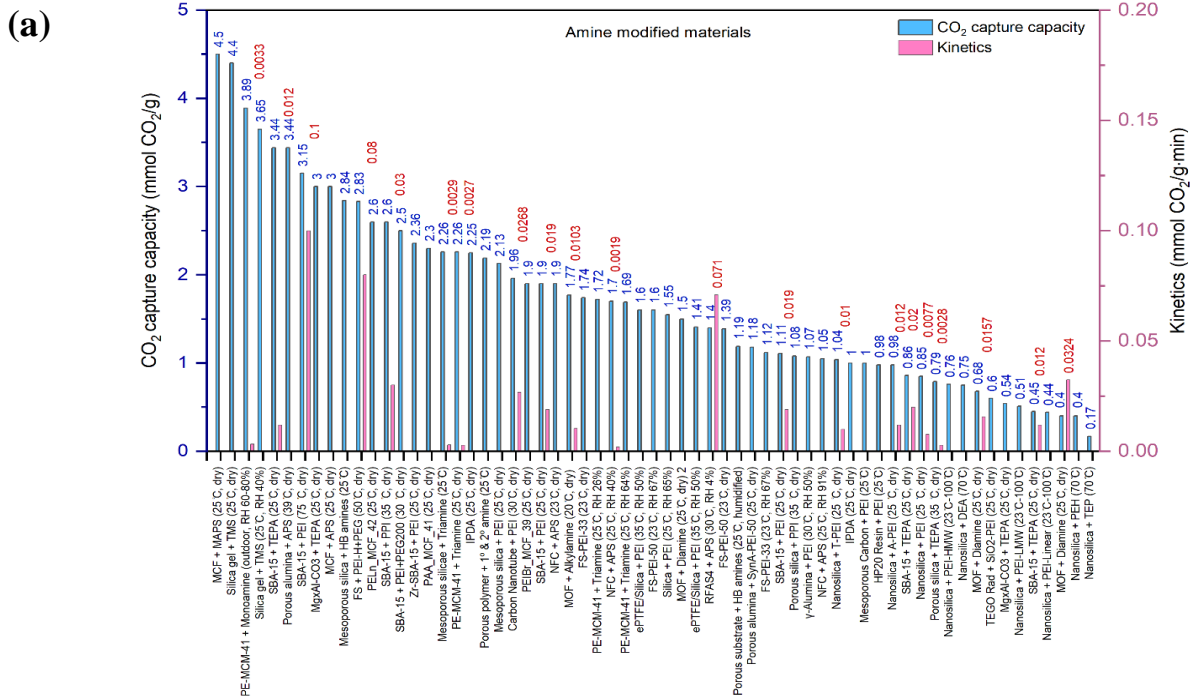

**(b)**

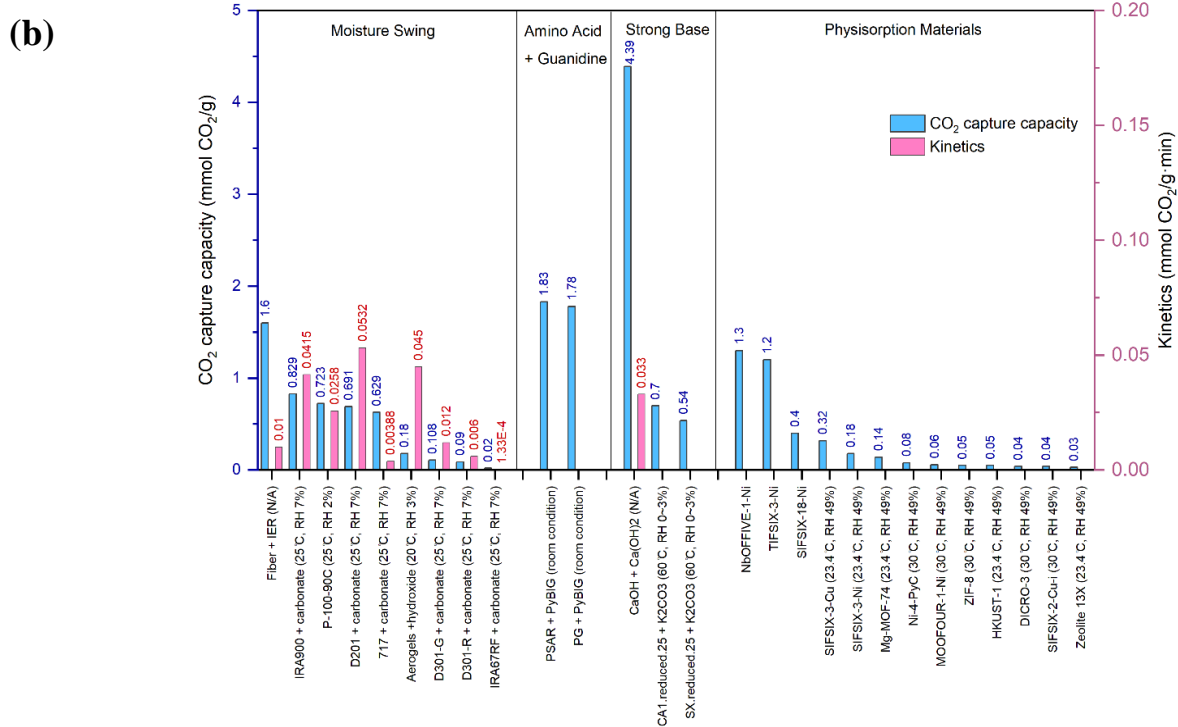

Figure 11. (a) $CO_2$ uptake capacities and kinetics of amine-functional materials, (b) $CO_2$ uptake capacities and kinetics of the technologies of moisture swing, amino acid+guanidine, strong base and physisorption.

## 3. The Importance of the Stability of $CO_2$ Sorbent

The development of a new class of thermally stable $CO_2$ capture materials is one of the main hurdles to efficient $CO_2$ collecting. Lackner's group[14,81,82,202] provided a techno-economic model consisting of a net-present value equation to determine the maximum affordable budget (MAB) of a sorbent for any DAC system based on factors that are well recognized. The MAB assessment of existing DAC sorbents used in previous research suggests that the high capacity, short cycle duration, and strong resistance to degradation of each sorbent are the three most important aspects for the economic viability of DAC systems. As new DAC sorbents are the primary focus of current research, it is strongly suggested that the stability of DAC sorbents be investigated. Table shows that for a lifetime of 100,000 cycles, MOF (Diamine) has the highest allowable budget of $180/kg because of high capacity and short loading time. Unfortunately, MOFs are among the most expensive sorbents with prices as high as $10,000/kg to $15,000/kg (and it actually cannot be used for 100,000 cycles). For a lifetime of 100,000 cycles, Ion Exchange Resin (IER) or chitosan has the allowable budget of $80-100/kg. The price is about $1-10/kg for IER and <$10-20/kg for chitosan. From the perspective of cost, the study expounds the importance of the stability of sorbents for the industrialization of DAC.

| Sorbent | Nominal Capacity ($C$) ($mol/kg$) | $C_{1/2}$* ($kg\ CO_2/kg\ Sorbent$) | $t_{cycle_{1/2}}$* ($min$) | Budget ($B$) ($\$/kg$) | | |
|---|---|---|---|---|---|---|
| | | | | 1k | 50k | 100k |
| TS-HAS-2.3 | 0.15 | 0.0033 | 105 | 0.1 | 4.32 | 6.96 |
| TS-HAS-2.9 | 0.23 | 0.0051 | 95 | 0.16 | 6.78 | 11.1 |
| TS-HAS-3.7 | 0.45 | 0.0099 | 140 | 0.32 | 12.04 | 18.24 |
| TS-HAS-5.4 | 1 | 0.022 | 163 | 0.72 | 25.52 | 37.26 |
| TS-HAS-8.4 | 1.4 | 0.0308 | 163 | 1.02 | 35.72 | 52.16 |
| TS-HAS-9.9 | 1.72 | 0.0378 | 167 | 1.26 | 43.52 | 63.18 |
| TS-PEI/silica | 2.36 | 0.0519 | 309 | 1.7 | 45.34 | 55.76 |
| TS-A-PEI/silica | 2.26 | 0.0497 | 196 | 1.64 | 53.9 | 75.12 |
| TS-T-PEI/silica | 2.19 | 0.0482 | 210 | 1.6 | 50.78 | 69.48 |
| TS-MOF (Diamine) | 2.83 | 0.0623 | 30** | 2.08 | 96.7 | 180.54 |
| MS-Chitosan | 1.68 | 0.0369 | 11 | 1.23 | 60.02 | 116.97 |
| MS-Carbon black | 0.14 | 0.0031 | 30** | 0.1 | 4.78 | 8.94 |
| MS-P-100-25C | 1.58 | 0.0348 | 30 | 1.16 | 53.98 | 100.8 |
| MS-P-100-90C | 1.58 | 0.0348 | 72 | 1.16 | 49.06 | 83.88 |
| MS-I-200-90C | 1.58 | 0.0348 | 108 | 1.16 | 45.3 | 72.4 |

Table: Sorbent budgets for lifetimes of 1,000, 50,000 and 100,000 cycles. The value of the sorbent (or the maximum allowed Budget is calculated based on the model described in the main text. In the evaluation, a discount rate r = 5%, and $P_{CO2}$ = \$50/t $CO_2$ have been assumed. The table is adjusted from Ref.[14]. *$tcycle$1/2, the time for half-loading the sorbent, is used as the cycle duration ($t$cycle) in the model and desorption is assumed to be virtually instant. For consistency we also used $C$1/2, which is half of the nominal capacity. **For an upper-bound sorbent budget the lowest value among the available data was used.

Amine sorbents own great capacity, but have degradation issues during the regeneration process under high temperature. The tiny, liquid amino molecules monoethanolamine (MEA), diethanolamine (DEA), and diisopropylamine (DIPA) have severe stability issues, due to their low boiling temperatures and high volatility[148]. Although PEIs with extremely large molecular weights (number-average molecular weights (Mn) tested up to 750 000) are relatively robust during $CO_2$ capture at high temperatures and in a vacuum, these structures do not support fast $CO_2$ diffusion into and out of the matrix. PEI with a high molecular weight might clog pores[109].

Pang *et al.* [203]showed the use of linear poly(propylenimine) (PPI) supported in silica as a sorbent for $CO_2$ removal from air. PPI-based sorbents for DAC are more effective than PEI-based sorbents because PPI is more resistant to oxidative degradation than PEI and therefore has a longer sorbent working life. Pang *et al*. [204]examined the tradeoff between sorption capacity and sorption-site accessibility across a variety of $CO_2$ collection circumstances. In comparison to linear PEI, linear PPI/SBA-15 composites preserved 65–83 percent of their $CO_2$ capacity following oxidative treatment, whereas linear PEI retained just 20–40 percent. In addition, the scientists proved the stability of linear PPI sorbents during 50 sorption/desorption cycles without any performance degradation. By incorporating a branching polyethylenimine (PEI) into a porous cross-linked poly(vinyl alcohol) (PVA) support, other researchers improved the material's resilience to degradation[205].

Wurzbacher et al.[168] first developed a cyclic temperature-vacuum swing (TVS) technique for separating CO2 from ambient air using a Class 2 sorbent. The cycle may function in both dry and moist air. In a solvent-free process, amine was grafted onto silica gel beads of 2–5 mm in diameter. After heating the sorbent material to temperatures ranging from 74 to 90 °C at a pressure of 150 mbar, the $CO_2$ taken from air with a relative humidity of 40% could be recovered. The sorbent exhibited outstanding stability through forty consecutive sorption/desorption cycles, with a cyclic $CO_2$ capture capacity of 0.8 mmol/g under dry circumstances and a regeneration chamber pressure of 200 mbar. Gebald et al. [169,206]discovered a new amine-based nanofibrillated cellulose (NFC)[207] sorbent. Over 100 consecutive cycles, the equilibrium $CO_2$ sorption capacity decreased by less than 5%, demonstrating the material's excellent stability.

Liquid-like nanoparticle organic hybrid materials (NOHMs) have demonstrated a high thermal stability. They are created by covalently or ionically grafting polymeric amine chains onto inorganic silica nanoparticles[183,208]. In-depth characterization studies have been carried out by Park's group to comprehend the material's characteristics and thermal stability[184-188]. This increase in oxidative thermal stability is attributable to the greater viscosity of the liquid-like NOHMs compared to the untethered polymer, and the bond stabilization of the ionically tethered polymer in the NOHM canopy[188]. Authors claimed that, due to their superior oxidative thermal stability, NOHMs can be used as functional materials for sustainable energy storage applications. To improve the highly viscous liquid-like bulk phase of NOHMs, Park's group developed hybrid $CO_2$ capture materials, solvent impregnated polymers (SIPs), that can enhance $CO_2$ capture kinetics and stability[189,209]. The incorporation and UV curing of liquid-like nanoparticle organic hybrid materials functionalized with polyethylenimine (NOHM-I-PEI) into a shell material yields gas-permeable solid sorbents with uniform NOHMs loading (NPEI-SIPs). Figure 12 shows the stability of SIPs over 20 consecutive $CO_2$ capture and release temperature swing cycles (50 °C to 100 °C or120 °C). At 100 and 120 °C, NPEI-SIP particles were sustained through 20 cycles with just a 2.6% loss.

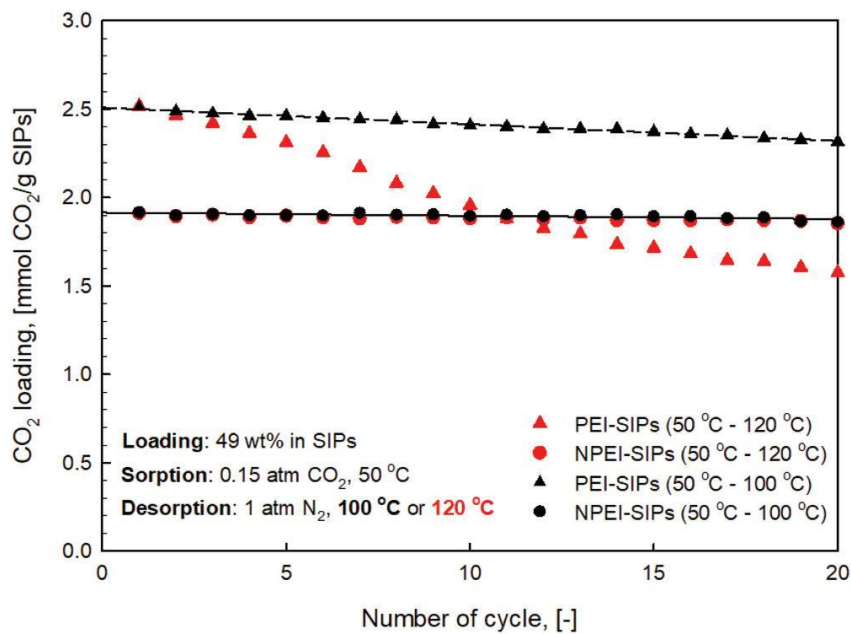

Figure 12. Over 20 CO₂ capture and sorbent regeneration cycles, the recyclability of NPEI-SIPs loaded with 49 wt% NOHM-I-PEI and PEI-SIP loaded with 49 wt% PEI was determined. In the CO2 capture stage at 50 °C, 0.15 atm dry CO2 balanced with N2 was employed, and the SIPs were regenerated under 1 atm N2 at 100 or 120 °C.

Still, there is a paucity of information in the literature on sorbent cycle length, stability, and lifespan. According to the available statistics, capacity reductions usually occur after less than 10 consecutive cycles. Our investigation reveals that air capture sorbents must go through tens of thousands of cycles[81]. Therefore, sorbent stability needs to be increased significantly.

## 4. Electrified Heating Technology for Sorbent Regeneration

It is feasible to significantly reduce greenhouse gas emissions through the development and implementation of electric industrial process heating systems[210]. A key advantage of electrified heating technology is the capacity to aim energy more accurately and, as a result, lower the total heat necessary to achieve the desired material or chemical change. Electrified heating technologies include resistive heating, induction heating, electromagnetic radiative heating, UV induced transformations, infrared heating, electron beam, plasma heating and transformations, *etc*. The regeneration procedure of CO₂ sorbent is necessary after it is fully loaded. The endothermic process of CO₂ desorption requires a high temperature of around 80 to 130 °C. Electrified heating technologies can be an alternative more efficient heating process compared to the combustion of fossil fuels.

## 4.1 Microwave Heating

Recently, the notion of employing microwave heating as an alternate method has been proposed for the regeneration of $CO_2$ sorbent. The effectiveness of MW irradiation depends on the MW energy's capacity to be absorbed by molecules having a dipole moment and transformed into heat[211], shown in Figure 13.

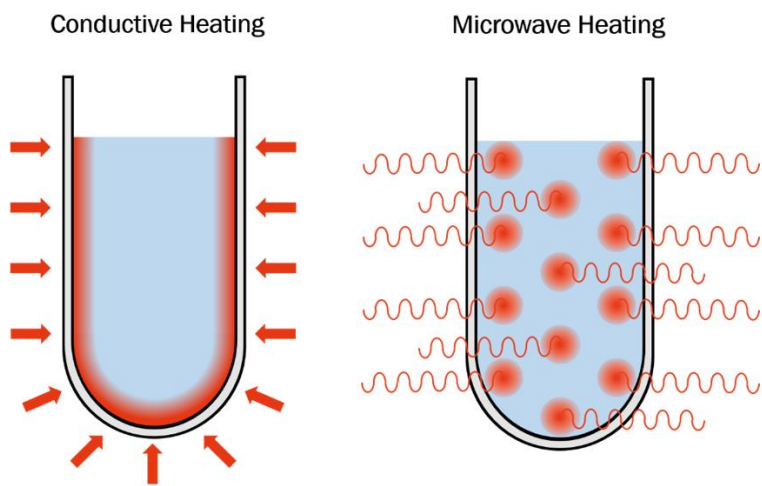

Figure 13. Approaches to heating: conductive heating and microwave heating.

The dielectric constant and dielectric loss factor are two important variables related to the MW method. A substance's capacity to retain energy is measured by its dielectric constant, and its capacity to transform electromagnetic energy into heat is calculated using its dielectric loss factor[212,213]. These two constants depend on the temperature, molecular interactions, and MW frequency in the solution[214]. MW has a reputation for being an instantaneous, volumetric heating process with an infinite capacity for heat transmission.

In comparison to traditional thermal desorption, Chronopoulus *et al*.[215] found that MW might potentially provide the option of an overall $CO_2$ desorption process that is four times faster. The MW for the regeneration process of amino-functionalized materials has also been observed approximately four times faster than conventional heating[216,217]. A research on the direct and indirect effects of microwaves on the Na-ETS-10 solvent was undertaken by Chudburry *et al*.[218].

After five cycles of $CO_2$ desorption, authors discovered that direct microwave regeneration produced a 22% greater desorption capacity and used 16.6% less energy than indirect microwave heating. McGurk stated that the MW could rapidly regenerate the MEA at low temperatures, which is about 70–90 °C, compared to the conventional thermal regeneration at high temperatures, 120–140 °C, which leads to the overall cost reduction. According to McGurk *et al.*[219], the MW could regenerate the MEA quickly at low temperatures around 70-90 °C, as opposed to the traditional thermal regeneration at high temperatures roughly 120-140 °C, which results in an overall cost savings. . Furthermore, Ji *et al.*[220] recently reported their work on microwave-assisted regeneration on a slurry of basic immobilized amine sorbent (BIAS) with 10 times accelerated $CO_2$ desorption and 48% lower electricity consumption compared to conventional thermal heating method, which authors argue was attributed to selective heating and increased polarity $CO_2$-adsorbed BIAS sorbents.

## 4.2 Induction Heating

Based on the well-known Néel and Brownian relaxation process, magnetic materials may produce heat on their own by converting electromagnetic energy into thermal heat when placed in an alternating magnetic field[221,222]. In an alternating magnetic field, Hill's group[223] employed a magnetic metal-organic framework (MOF) adsorbent to quickly release the adsorbed $CO_2$ by rapidly raising the temperature inside the magnetic adsorbent. When an alternating magnetic field is applied to a MOF mother liquid containing carboxylic acid-decorated MNPs, each MNP "nanoheater" generates heat locally, resulting in a solvothermal microenvironment (Figure 14). Furthermore, the MNPs serve as a nucleation surface for MOF development. It was therefore suggested to use a method known as magnetic induction swing adsorption (MISA). They then achieved quick $CO_2$ desorption at the magnetic field with energy efficiency of no more than 60% by using magnetic MgFe2O4@UiO-66[224].

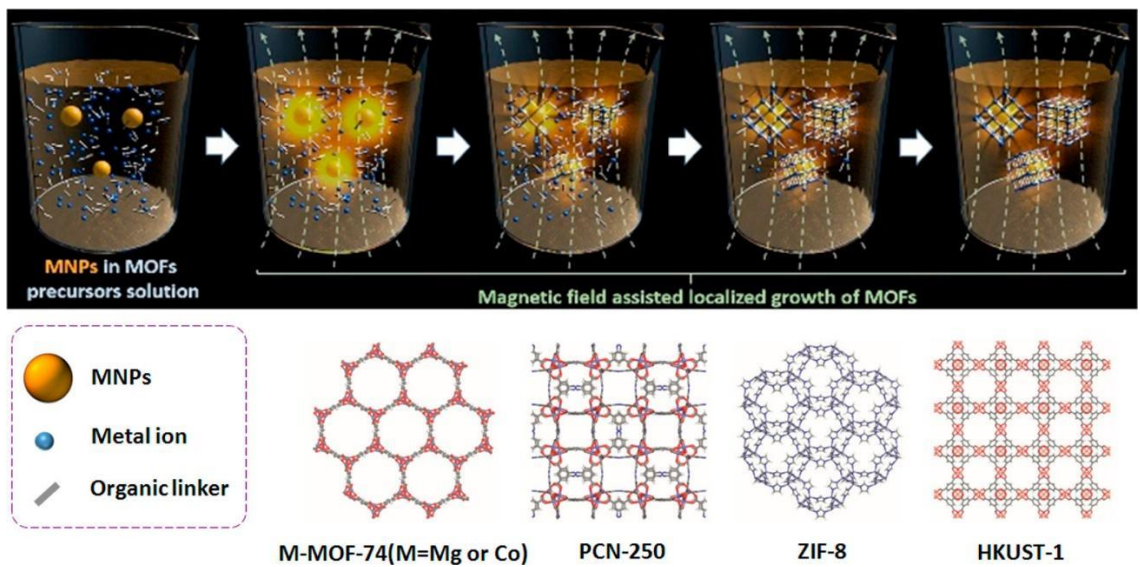

Figure 14. Diagram illustrating MOF development in a MIFS method, as well as crystal structures of MOFs synthesized with MIFS.

By synthesizing magnetic nanoparticles on NPC support, Lin *et al.*[225] created Fe3O4/nitrogen-doped porous carbon (NPC) sorbents. In the meantime, a $CO_2$-TSA in situ electromagnetic induction heating (EMIH) device was developed to assess the performance of the $CO_2$-TSA in situ electromagnetic induction heating (Figure 15) and to further shed light on the connection between $CO_2$ desorption rate and energy effectiveness. The best adsorbent was determined to be $Fe_3O_4$/NPC-15, which had a $CO_2$ capacity of 2.64 mmol g-1 and saturation magnetization of 15.51 emu g-1. Upon optimization, the fixed target temperature heating mode at 110 °C exhibited the best regeneration performance with the desorption rate of 3.27 mg g–1 s–1 and the energy efficiency of 79.2%, which are not only superior to those of the reported MISA technology but also considerably more efficient than the typical convective-heat-transfer TSA performance. Researchers also utilized EMIH for $CO_2$ desorption on the sorbents of 13X zeolite and $Fe_3O_4$[226], as well as $Fe_3O_4$@HKUST-1[227].

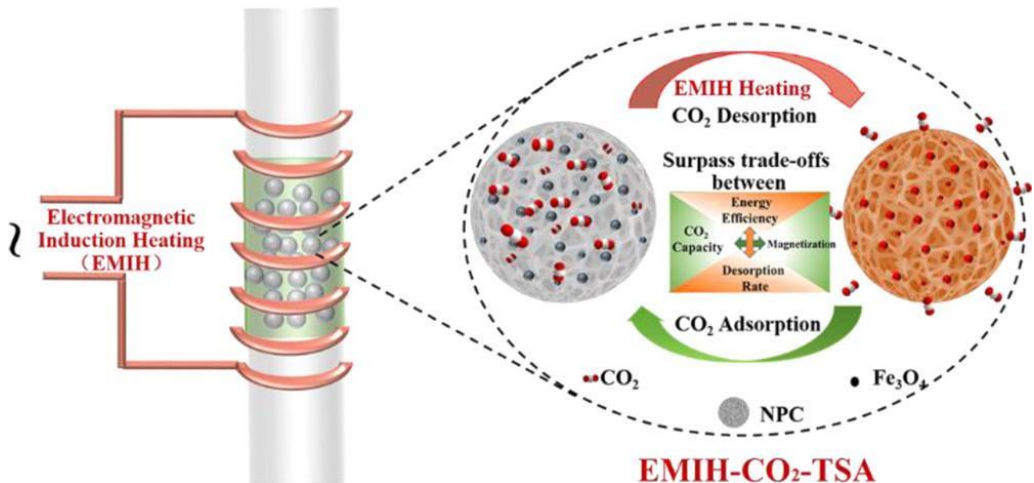

Figure 15. Electromagnetic induction heating for Temperature swing adsorption (TSA)-based $CO_2$ capture (EMIH-$CO_2$-TSA).

## 4.3 Plasma-assisted $CO_2$ desorption

Plasma technology is another non-thermal method explored to regenerate $CO_2$ capture materials. According to Li *et al.* [228], plasma was used to control the desorption process instantly. Turning on plasma ignition induced the desorption of $CO_2$ from the hydrotalcite surface soon after, and as the plasma was switched off, the desorption process stopped immediately. At the same time, the plasma treatment also split $CO_2$ and achieved an average conversion of 41.14% to CO. Authors utilized the plasma technology to merge both desorption of $CO_2$ and conversion and claimed that the technology could reduce the cost and complexity of the DAC. However, according to this research, a considerable improvement in energy efficiency must be made for plasma-assisted $CO_2$ capture and conversion to be applicable.

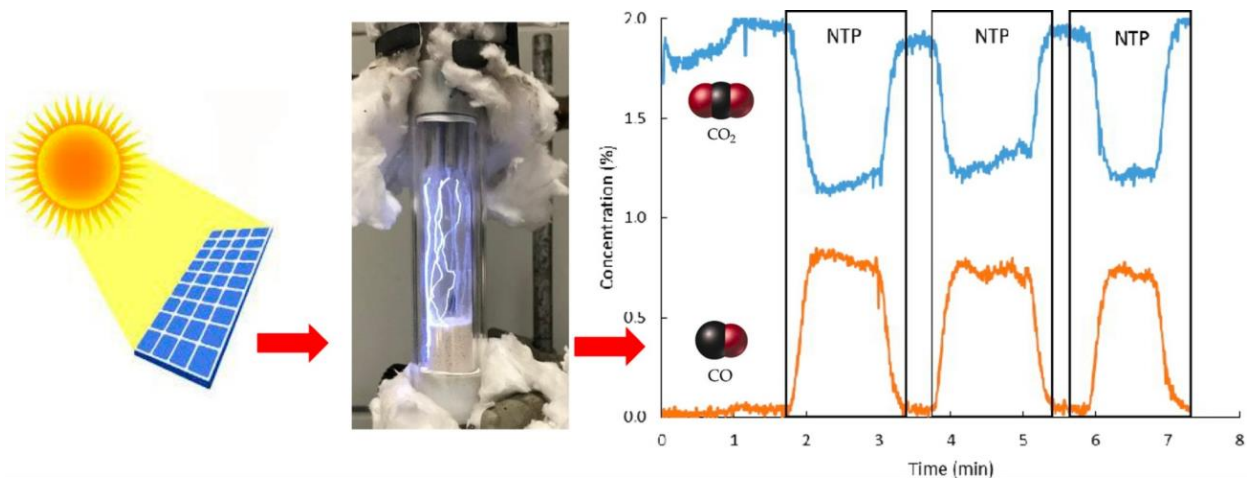

Figure 16. Non-thermal plasma induced $CO_2$ reduction via photovoltaic energy.

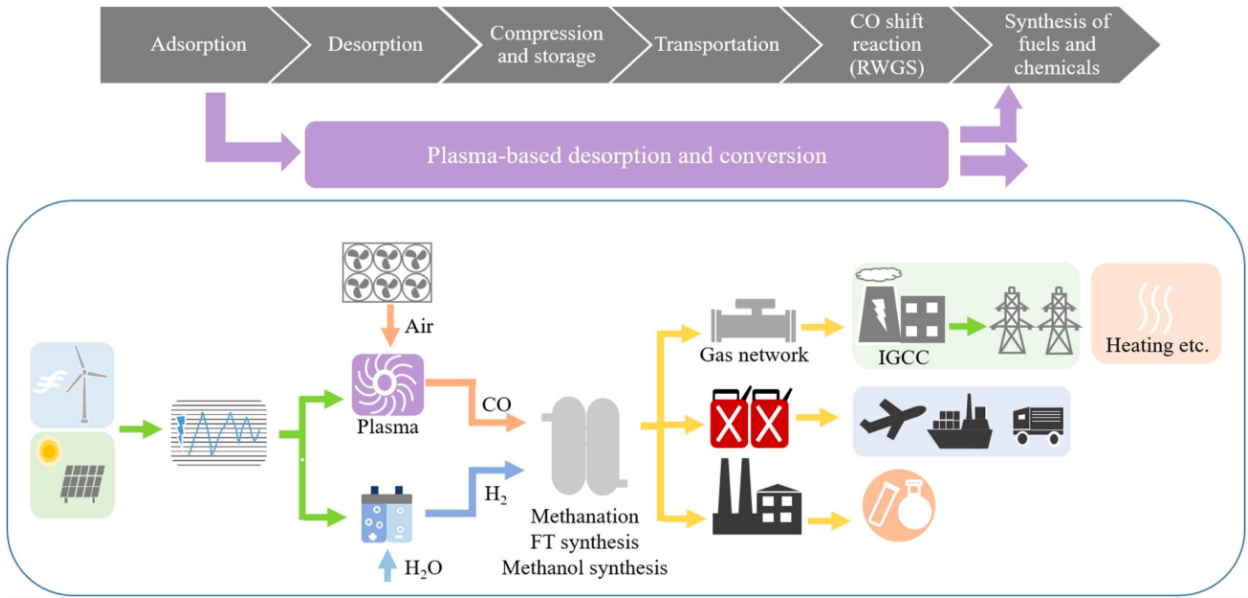

Figure 17. Schematic diagram of syngas production via Direct Air Capture of $CO_2$ and conversion by plasma.

Pou *et al.* [229] explained that non-thermal plasma produced with the photovoltaic system could generate the optimum electron energy to break the chemical bonds at room temperature and under atmospheric pressure, which makes the desorption process more energy efficient than thermal methods. They calculated the $CO_2$ emissions compensated value using photovoltaic electricity to be 67% higher than that of using an electricity mix, proving that coupling non-thermal plasma with a photovoltaic system considerably boosts sustainability.

## 5. Sub-ambient Temperature Surrounding for DAC

To maximize the effectiveness of DAC processes, it is crucial to understand how well DAC materials absorb $CO_2$ in actual operating conditions. Unfortunately, little is known about capture effectiveness of these materials at various operating temperatures. The investigation of sorbent performance was limited to temperatures that were both ambient and above ambient (20 °C).

Due to the fact that more than 80% of the world's land has an annual average temperature below 25 °C, research on DAC under sub-ambient settings has the ability to enhance DAC technology development and the possibility for general deployment on a global scale (Figure 18)

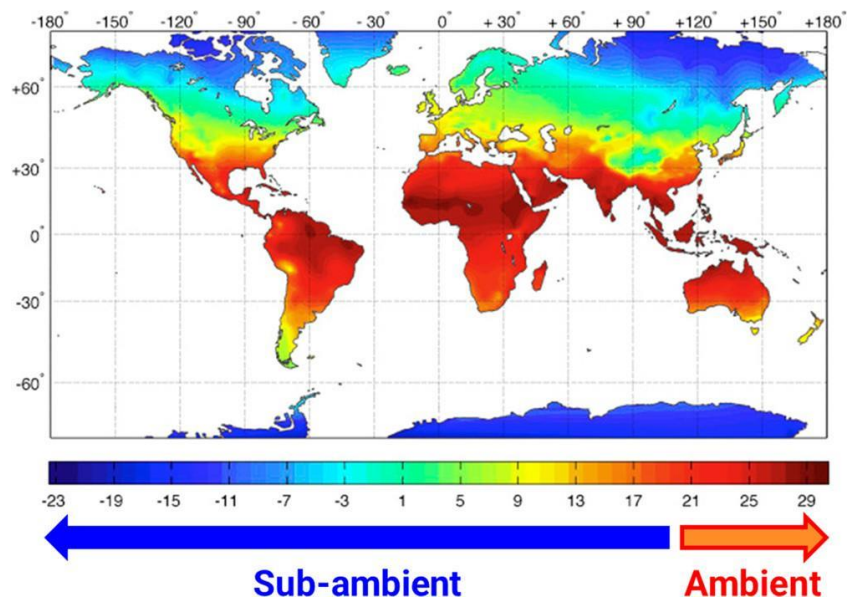

Figure 18. Map of world average temperature (°C)

At sub-ambient temperatures (particularly 0 °C), it is anticipated that the physical and chemical properties of many CO2 sorbents will undergo substantial modifications. Therefore, sorbents optimized for ambient DAC must be reevaluated and, in many instances, modified in order to attain optimal performance at sub-ambient circumstances. DAC may provide some benefits in sub-ambient temperature circumstances. The energy required to desorb water across temperature or temperature-vacuum swing adsorption cycles may be limited by the lower absolute humidity at colder temperatures. Additionally, the use of physisorbents rather than chemisorbents may be made possible by freezing temperatures.

Miao *et al.*[230] studied how the DAC performance of the polyamine-loaded mesoporous silica was affected by operating temperatures. Researchers found that at lower adsorption temperatures below 15 °C, the $CO_2$ adsorption selectivity of polyethylenimine-loaded mesoporous silica reduces. The ideal desorption temperatures, which range from 90 to 120 °C, as well as the adsorption kinetics and long-term stability were also parametrically examined by the authors. Figure 19 shows that the optimal adsorption temperatures for 50% PEI/SBA-15 working at simulated atmospheric conditions are 45 °C. $CO_2$ capture kinetics reduces with the decrease of temperature from 45 °C.

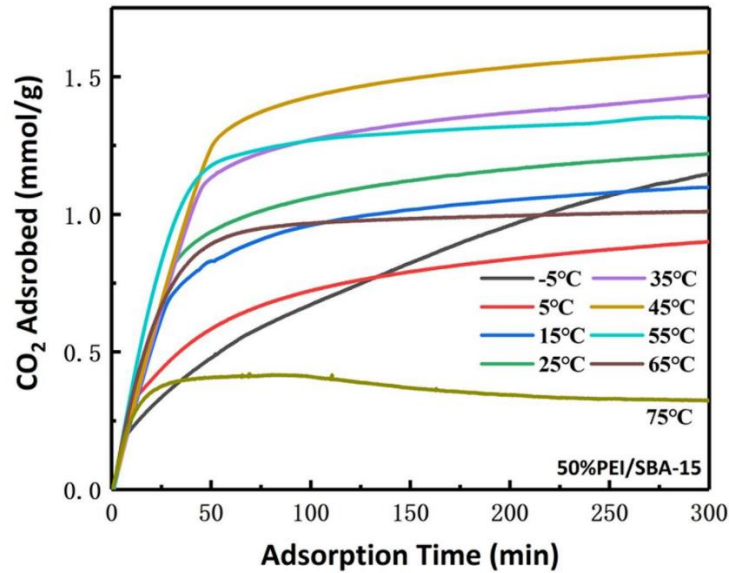

Figure 19. Comparison of the capacity of 50% PEI/SBA-15 at different adsorption temperatures.

According to Rim *et al.*[231], amine-impregnated MIL-101(Cr) materials (PEI, poly(ethylenimine), or TEPA, tetraethylenepentamine), provide promising adsorption and desorption behavior under DAC circumstances in both the presence and absence of humidity throughout a wide range of temperatures (20 to 25 °C). The sorbents exhibit varying $CO_2$ capture behavior depending on the amine loading and adsorption temperature. The sorbents exhibit weak and robust chemisorption-dominant $CO_2$ capture behaviors, respectively, with amine loadings of 30 and 50 weight percent. Interestingly, due to enhanced weak chemisorption at 20 °C, the $CO_2$ adsorption capacity of 30 wt% TEPA-impregnated MIL-101(Cr) rose noticeably from 0.39 mmol/g at ambient conditions (25 °C) to 1.12 mmol/g. The sorbents also demonstrate promising working capacity (0.72 mmol/g) across 15 cycles of minor temperature swings with extremely low regeneration temperatures (20 °C sorption to 25 °C desorption). Under humid conditions, the sorbents' sub-ambient DAC performance is improved even more, displaying promising and steady $CO_2$ working capabilities over several humid small temperature swing cycles. These findings show that properly engineered DAC sorbents are capable of mild chemisorption at low temperatures, even in the presence of humidity. By using the modest temperature changes made possible by this weak chemisorption behavior, significant energy savings may be achieved. According to this research, further studies are needed on DAC materials that can function at low, sub-ambient temperatures in preparation for potential deployment in temperate and polar climes.

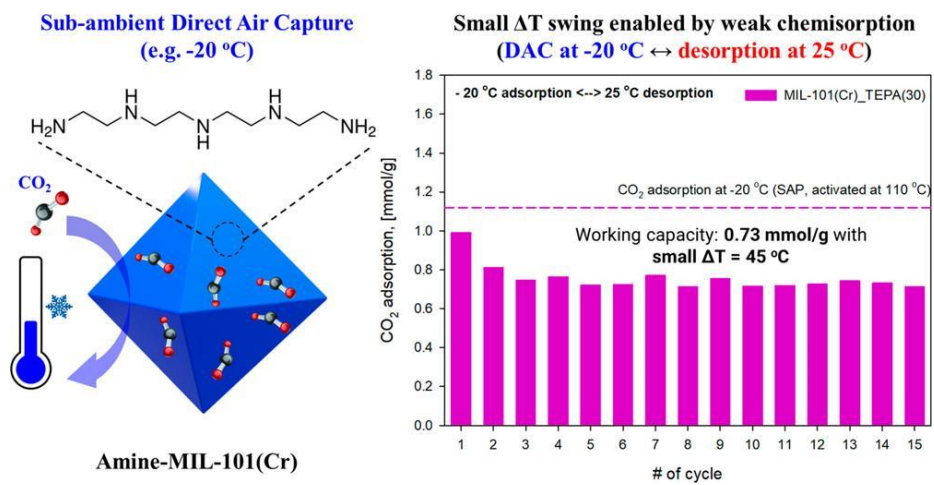

Figure 20. Capacity of sorbents with an ultralow regeneration temperature (−20 °C sorption to 25 °C desorption).

## 6. Computational Modeling

Computational approaches are important and complementary adjunct to experiments for the purpose of comprehending the processes of sorbent-$CO_2$ reactions, developing carbon capture agents with acceptable costs and toxicity, as well as designing Direct Air Capture processes. Quantum chemistry (QC) may accurately and efficiently predict electronic and structural features, including reaction energetics and free energies of reaction, transition state energies, and structures of reaction species[232-239]. Classical molecular dynamics (MD) and Monte Carlo (MC) modeling employing molecular mechanics force fields also serve a crucial role in relating experimental data of macroscopic scale processes to the molecular level. For instance, they can develop a fundamental understanding of the impact of solute–solvent interactions and the specifics of reaction dynamics, transport behavior (e.g., diffusivity, viscosity, and mass transfer) and other physicochemical phenomena on the performance of solution $CO_2$ capture and release[240-247]. The thriving machine learning (ML) has demonstrated considerable potential for expediting the development of porous materials for carbon capture[248,249]. Multiphysics modeling coupled with fluid dynamics, heat and mass transfer, and chemical reaction may successfully design a $CO_2$ capture filter and estimate the performance of a scaled-up carbon capture system with a specified degree of confidence[250-252]. The combination between chemical engineering experiments and computational modeling can give an efficient and cost-effective method of interpreting data at the pilot-plant scale, thereby leading research into process alternatives and optimizing the deployment of a plant.

## 6.1 Mechanism Study Based on Quantum Chemistry

Two alternative sorption processes are found for Amine-functional sorbents: 1) Under dry circumstances, primary and secondary amines combine with $CO_2$ to produce carbamate or carbamic acid. 2) Amines react with $CO_2$ in the presence of moisture to create bicarbonate. However, the processes of $CO_2$ sorption remain unknown, and various groups typically ascribe conflicting infrared bands to bicarbonate. This raises questions about the presence of bicarbonate.

da Silva $et$ $al.$[235,236] pioneered the mechanism study of amine sorbents by atomistic modeling. They employed ab initio calculations to determine the most likely process of carbamate production in alkanolamines interacting with $CO_2$. A zwitterion intermediate with a considerable lifespan in the system appears implausible. The experimental findings appear to be in excellent accord with a single-step process. Using the coordinate driving method, a series of computations were carried out in which the response coordinate was systematically modified while the other geometric parameters were optimized. There was no intrinsic energy barrier in the subsequent proton transfer from the zwitterion to another MEA molecule to create carbamate. The zwitterion was not a local minimum in the gas phase, but it became one when solvent was added. The study demonstrated that the factor of solvent was crucial for the stability of zwitterions. Therefore, the solvent impact must be meticulously accounted for while analyzing the $CO_2$ and amine interaction.

The study result from Arstad $et$ $al.$[237] is in contrast to those from da Silva $et$ $al.$[235,236] and Shim $et$ $al.$[238] which reveal that there is no reaction energy barrier in the proton abstraction mechanism that forms carbamate. By combining DFT calculations for geometry optimization and energy corrections for electron correlation, the quantum description of molecules in the gas phase was achieved[237]. Arstad $et$ $al.$ postulated a two-step process based on an intermediate carbamic acid. Additional water or amine molecules moved a proton from the nitrogen atom of the amine to a terminal CO2 group in the zwitterionic encounter complex $RNH_2^+CO_2^-$, according to their calculations. As indicated in Figure 21, at the G3MP2B3 level of theory, the energy barrier for the proton migration mechanism from the zwitterion to the direct catalyst MEA was 9.3 kcal/mol. Recent DFT calculations suspect the existence of zwitterions on solid amines again[233]. In order to clarify the optimal $CO_2$ adsorption method and deactivation process, Buijs $et$ $al.$[253-256] carried out a series of atomistic modeling investigations. They proposed that amine or amine-$H_2O$ catalyzed reactions are the main mechanisms through which carbamic acid mostly formed in the presence of $CO_2$ and $H_2O$.

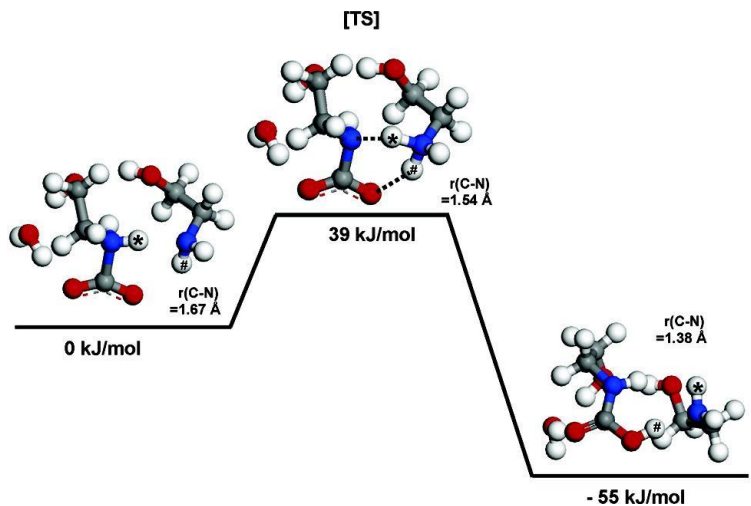

Figure 21. Structures and relative energies to the start complex of the reaction between $CO_2$, monoethanolamine (MEA) catalyzed by another MEA, and water.

Shi *et al.*[129] demonstrate, by means of quantum modeling, a sequence of unorthodox chemical processes in which the degree of hydrolysis of basic salts containing multiple water molecules is markedly different from that in bulk water and can be controlled by adjusting the relative water content. Figure 22 shows the free energy change of the reaction between basic salts and different number of water molecules. This discovery has been applied to the design of efficient absorbents for $CO_2$ capture by controlling the amount of water molecules.

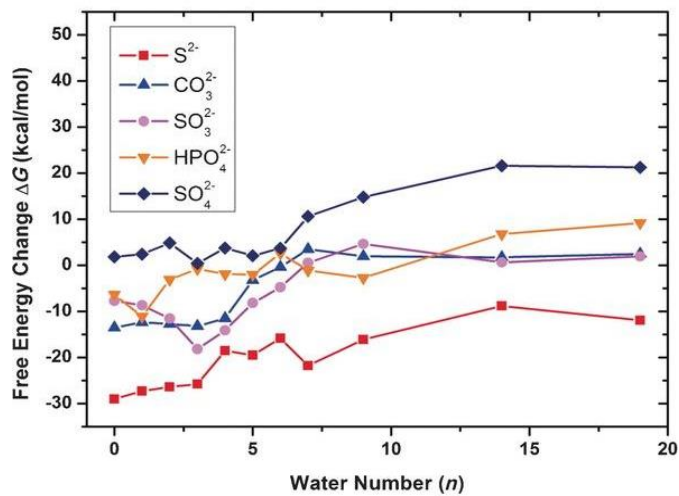

Figure 22. Free energy changes of the chemical reaction, $X^{2-} + nH_2O = HX^- + OH^- + (n\text{-}1)H_2O$. X is basic ion, $n$ is water number.

## 6.2 Mechanism Study Based on Molecular Dynamics

To appreciate reaction processes and the behavior of $CO_2$ and sorbents, it is vital to comprehend the structure of participant molecules and the interactions that occur in both dynamic and equilibrium phases. These functionalities can be clarified using the MD molecular simulation approaches. While abundant experimental data on $CO_2$ capture, it remains difficult to model these systems effectively enough to recreate, for example, temperatures of reaction that are consistent with experimental results[257,258]. To study liquid-solid-$CO_2$ systems, sophisticated molecular simulation techniques have been developed. Incorporating density functional theory (DFT) into molecular dynamics is one example[259].

A methodology combining MD and DFT, shown in Figure 23, was utilized by Shi *et al.*[192,260] to calculate energy states connecting aqueous states to ionic states in the vacuum during $CO_2$ capture and release process. The following studies analyzed the effects of parameters of sorbents on $CO_2$ capture efficiency and lead the way toward the optimization of sorbents for DAC[86,132].

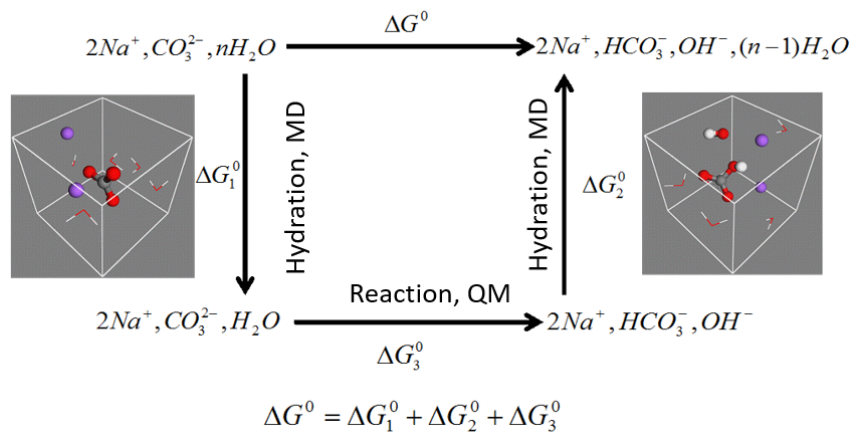

Figure 23. Thermodynamic cycle for calculating reaction energy change with water numbers.

Iida *et al.*[261,262] investigated the effect of solvent water on the formation of amine-$CO_2$ bonds using RISM-SCF-SDD[263,264], a hybrid technique combining QC for the solute and statistical mechanics for the solvent. It is acknowledged as a substitute for the QM/MM approach. They found that the barrier to reaction and the stability of the transition state in aqueous solution depended on an interaction between the hydration and dehydration of O and N. There was a single minimum at a reaction coordinate distance of 3.0 with no barrier present in the gas phase. In good agreement with the experiment, a stable structure was observed in aqueous solution at a reaction coordinate distance of 1.6, with a transition state at 2.0 and an activation enthalpy of 9.3 kcal/mol, after the formation of identical intermediates in the gas phase.

Sumon *et al.*[246] investigated the suggested zwitterion, termolecular, and carbamic acid processes using the semicontinuum (cluster plus continuum) solvation model while considering proton relays. They intended to remedy the poor stability forecast of aqueous carbamates that has persisted for decades. The chemical pathways for a single Monoethanol-amine (MEA) and two MEAs were computed using a solvation model with varying quantities of water molecules. The simulations of the single MEA route utilized different number of water molecules. The two MEA models were investigated not only because the abstracting base in the proposed mechanisms could be a second MEA molecule, but also because $CO_2$ capture experiments demonstrate that the proton released by the zwitterion (even if released initially to a bulk water molecule) migrates rapidly to a second MEA molecule, inhibiting the decomposition of carbamate back to $CO_2$ + MEA. For accurate modeling of chemical pathways, it has been demonstrated that the insertion of numerous explicit water molecules is essential.

### 6.3 Discovery of materials for carbon capture by machine learning

The thriving field of machine learning (ML) has been brought to several fields of materials science[265-268], which has demonstrated considerable potential for expediting the development of materials for carbon capture[269]. Farmahini and coworkers[248] have provided a summary of the recent development of multiscale and performance-based screening workflows for use in post-combustion carbon capture. Yan *et al.*[249] demonstrated how ML has been utilized to enhance the carbon capture, transport, usage, and storage value chain and made recommendations for further research.

Dai's group trained deep neural networks (DNNs) to estimate the $CO_2$ adsorption capacity using textural parameters as input features, including specific surface area, micropore volume, and mesopore volume[270,271]. First, it was discovered that only when three textural characteristics are employed for model training is the highest output prediction accuracy attained. In order to retrain the model, temperature (T) and pressure (P) were added to the input characteristics. They discovered that when pressure rose, the contribution from the mesopore increased, which is consistent with other studies[272].

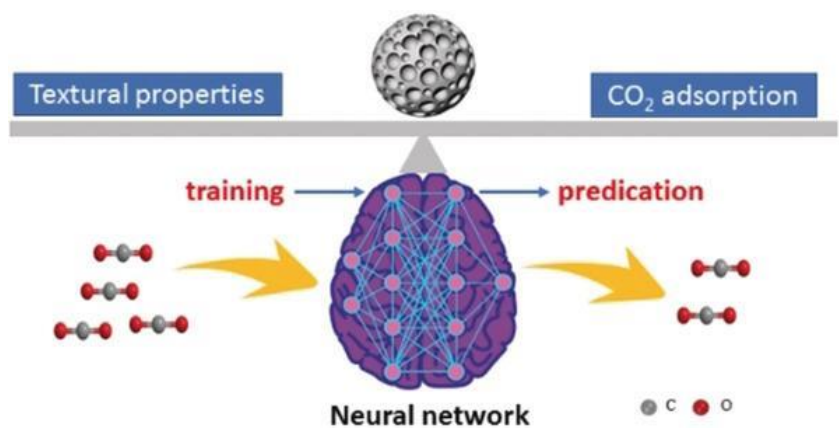

Figure 24. The prediction of $CO_2$ sorption of porous materials.

Application of machine learning in MOFs for carbon capture seeks to develop structure–performance connections and pick the most accurate descriptors for predicting the $CO_2$ adsorption capacity, working capacity, and selectivity[273-278]. Woo *et al.*[274] created a quick and precise ML model for screening MOFs with increased $CO_2$ adsorption capability. Later, they increased the prediction accuracy rate, identifying 994 of the top 1000 MOFs from a test set of 70,000 MOFs[275]. Boyd *et al.* [273] designed and prepared MOFs with high $CO_2$ adsorption capability for wet flue gas using a data-driven methodology. They discovered that MOFs with parallel aromatic rings had almost optimal interactions with $CO_2$ and a low Henry coefficient for water. The $CO_2/N_2$ selectivity investigation revealed that the pore structure of MOFs contributed to a high $CO_2$ uptake at high relative humidity by preventing the formation of hydrogen bonds.

## 7. Dual-functional Materials for $CO_2$ Capture and Conversion

In addition to efficient absorption of $CO_2$ from the atmosphere, usage of the absorbed $CO_2$ is a crucial feature in combating global climate change caused by excessive emission of green gases. Several $CO_2$ conversion technologies, including electrochemical[279-282] thermochemical[283-286], as well as photochemical and photoelectrochemical[287-289] methods, are in the process of development shown in Figure 25.

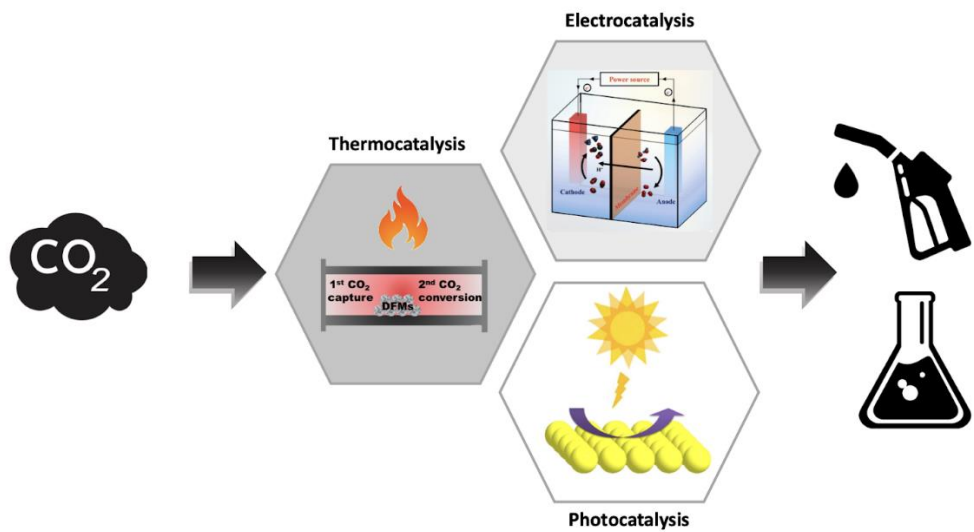

Figure 25. Three technologies of $CO_2$ conversion by using dual functional materials (DMF)

By combining a $CO_2$ capture phase with a $CO_2$ conversion step using dual-functional materials (DFM), it is possible to deploy a reactive carbon capture strategy. In this concept, once the $CO_2$ capture materials bind to $CO_2$, they do not release $CO_2$ via an energy-intensive desorption process, but instead act as an electrolyte additive to transport the captured $CO_2$ molecules to the catalytic sites for the subsequent thermochemical or electrochemical $CO_2$ reduction. The $CO_2$ capture materials are renewed for the subsequent $CO_2$ capture cycle following the conversion phase. Thus, the overall energy need for $CO_2$ capture and conversion might be drastically decreased. For example, A combination of $CO_2$ adsorbents (i.e. CaO) and common active metals for $CO_2$ conversion (i.e. NiO and FeOX) may be used to create catalytic systems such as Ni/CaO and $Fe_2O_3$/CaO-$Al_2O_3$, which demonstrate high $CO_2$ adsorption capacity and product variety by creating CO, $CH_4$, and methanol[290,291]. These characteristics enable DFM to first absorb $CO_2$ from diluted flue gases and then sequentially complete the catalytic conversion process when the input gas changes from flue gas to reducing agent[292].

Modern DFM materials with dual roles for $CO_2$ adsorption and chemical conversion stages include metal-organic frameworks (MOFs), covalent organic frameworks (COFs), and 2D transition metal carbides and nitrides (MXene)[293-296]. The majority of these DFMs however still have issues with low $CO_2$ selectivity, poor electrical conductivity, poor chemical stability, harsh/complicated chemical production, high cost, *etc*. which severely limits the overall efficiency of the process[297,298]. As a result, research is increasingly focusing on the development of innovative DFMs for increased $CO_2$ collection capacity and improved conversion properties (i.e., activity and stability).

## 7.1 Electrochemical $CO_2$ reduction

Electrochemical $CO_2$ reduction is frequently one of the most effective ways to reduce the excessive $CO_2$ output because of the following advantages: The electrocatalytic process uses electrons rather than hydrogen to reduce; it takes place under mild circumstances; the products may be reformed utilizing a variety of electrolyte potentials, temperatures, and concentrations; and it can be controlled to minimize side products[299]. Numerous substances, such as liquid products like formic acid (HCOOH), methanol ($CH_3OH$), and ethanol ($C_2H_5OH$), as well as gases like carbon monoxide (CO), methane ($CH_4$), and ethylene ($C_2H_4$), may be successfully separated from process byproducts using a condenser.

To keep the electrodes adequately moist and prevent solvent loss, it is preferable for the electrolytes to have low volatility. Electrolytes with high salt concentrations, low volatility, and customizable physical characteristics include ionic liquids (ILs)[300,301] and deep eutectic solvents (DESs)[302-304]. As a result, they are potential multifunctional electrolytes for processes that combine capture and conversion. However, the high viscosities and poor conductivities that these electrolytes often exhibit result in transport rates that are substantially slower than ideal for practical applications. It could be interesting to construct these electrolytes using aqueous systems[305].

Novel materials, such as nanoparticle organic hybrid materials and ionic liquids, have been proposed as promising options for reactive carbon capture due to their high $CO_2$ capture capabilities and charged nature. They may be constructed by optimizing the $CO_2$ binding energy. Research has demonstrated that these innovative electrolyte materials can enhance $CO_2$ conversion rates and product distributions[306]. For instance, Park's group demonstrated that the reactivity of nanoparticle organic hybrid materials attached to $CO_2$ differs from the reactivity of free $CO_2$, demonstrating the possible co-catalytic function of nanoparticle organic hybrid materials during electrochemical $CO_2$ conversion[307-312]. The schematic of an electrochemical cell containing NOHM-based electrolytes has been shown in Figure 26.

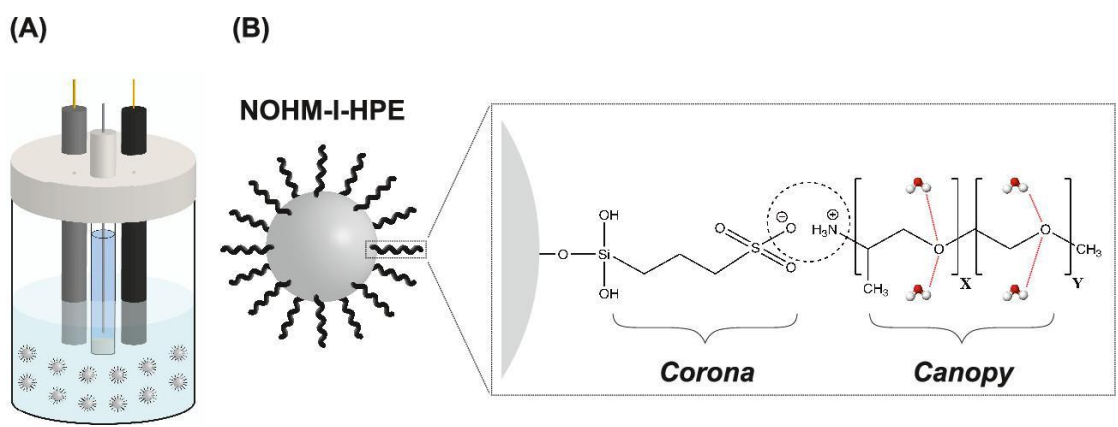

Figure 26. (A) Schematic of an electrochemical cell with electrolytes based on NOHM. (B) The structure of the NOHM-I-HPE synthesized for this work. Illustrated are potential hydrogen bonding interactions with the secondary fluid (i.e., water).

As of this writing, the solid oxide electrolysis cell (SOEC) is the special technology that brings $CO_2$ reduction to CO closer to commercialization[313,314], shown in Figure 27. Because the ionic conductivity of electrolyte materials increases significantly with temperature, pure $CO_2$ or a mixed gas with high concentration of $CO_2$ must be added to SOECs. Normally, SOECs are configured to operate at a temperature of between 600 and 850 °C[314]. In order to attain zero emissions, low-temperature electrolysis is better suited for removing $CO_2$ from the air and decreasing it without having to deal with $N_2$, $O_2$, and $H_2O$.

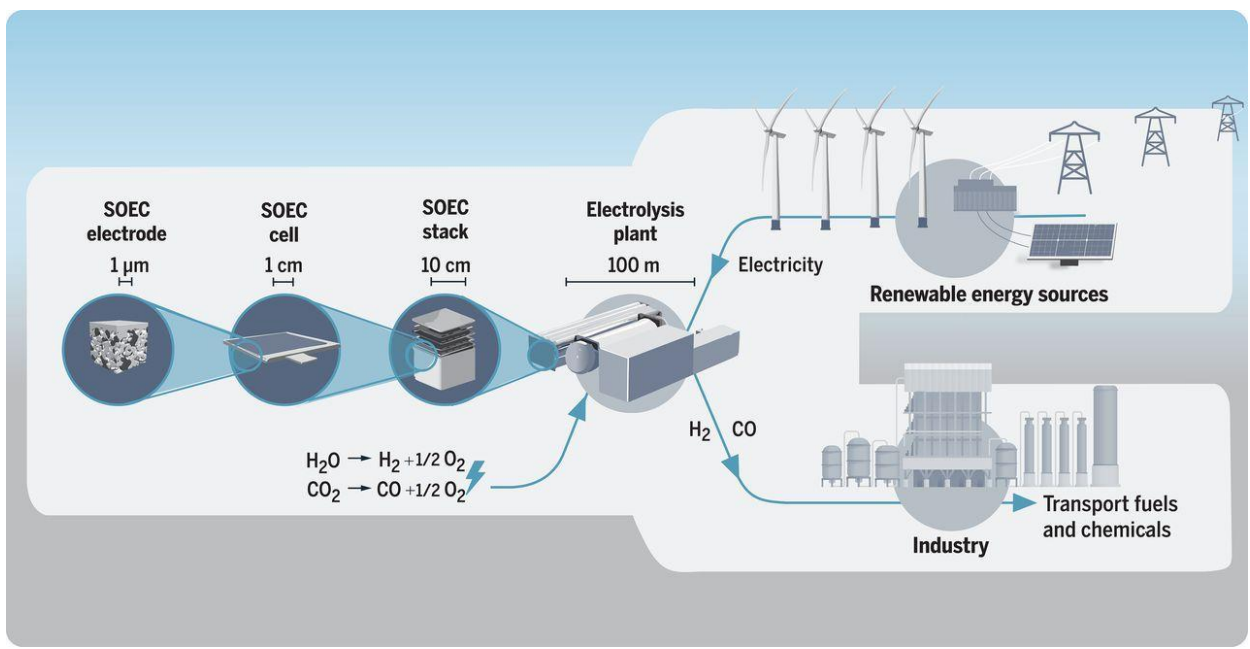

Figure 27. At the electrodes of a solid oxide electrolysis cell (SOEC), water or carbon dioxide are separated. SOEC stacks, which are composed of several cells, are then composed into SOEC plants. The manufacturing of chemicals and transportation fuels may be separated from the usage of fossil fuels when renewable power is employed. Other electrolysis methods cannot match the electrolysis efficiency of SOECs, which run at high temperatures.

## 7.2 Thermochemical $CO_2$ reduction

Recent investigations have revealed that dual function materials (DFM) containing noble metals such as Ni[315-321] or Ru[322-335] supported by r-$Al_2O_3$ integrate adsorption and conversion to methane and other light alkanes[336-342]. equations (1) - (5) and Figure 28 depict the topography inside the Ru-based DFM. $CO_2$ was collected at the interface between $Na_2O$ and oxygen to create carbonated species, whereas Ru was oxidized to generate $RuO_2$ or $RuO_x$. $RuO_x$ would be reduced to Ru after the introduction of hydrogen, allowing $CO_2$ molecules to transfer from $Na_2O$ sites to Ru sites for the methanation process.

$$Na_2O + CO_2 \rightarrow Na_2CO_3 \#(1)$$

$$Ru + O_2 \rightarrow RuO_2(RuO_x)\#(2)$$

$$RuO_2(RuO_x) + 2H_2 \rightarrow Ru + 2H_2O \#(3)$$

$$Na_2CO_3 + Ru \rightarrow Na_2O + Ru - CO_2 \#(4)$$

$$Ru - CO_2 + 4H_2 \rightarrow CH_4 + 2H_2O + Ru \#(5)$$

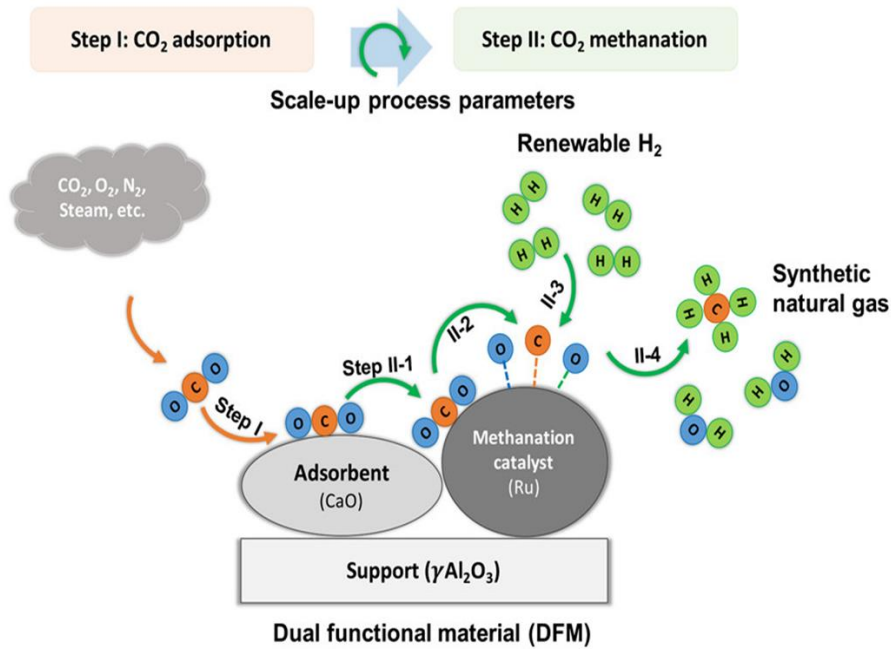

Figure 28. Adsorption and Methanation of $CO_2$ with dual functional catalytic materials.

The results of 10 cycles of adsorption, desorption and $CH_4$ production in dry air condition are shown in Figure 29a and 5 cycles in 2 % wet air condition shown in Figure 29b. The average adsorption under humid condition is about 1.3 $mmol_{CO2}/g_{DFM}$, which is 2.36-fold increasing compared to dry condition. It is extraordinarily desired to convert $CO_2$ directly into methane to decrease environmental $CO_2$ emissions and store renewable energy[343]. In addition, the effect of Rh particle size on the activity and reaction process of $CO_2$ methanation over Rh/TiO$_2$ catalysts at low temperatures (85 - 165 °C) and atmospheric pressure[344] was examined.

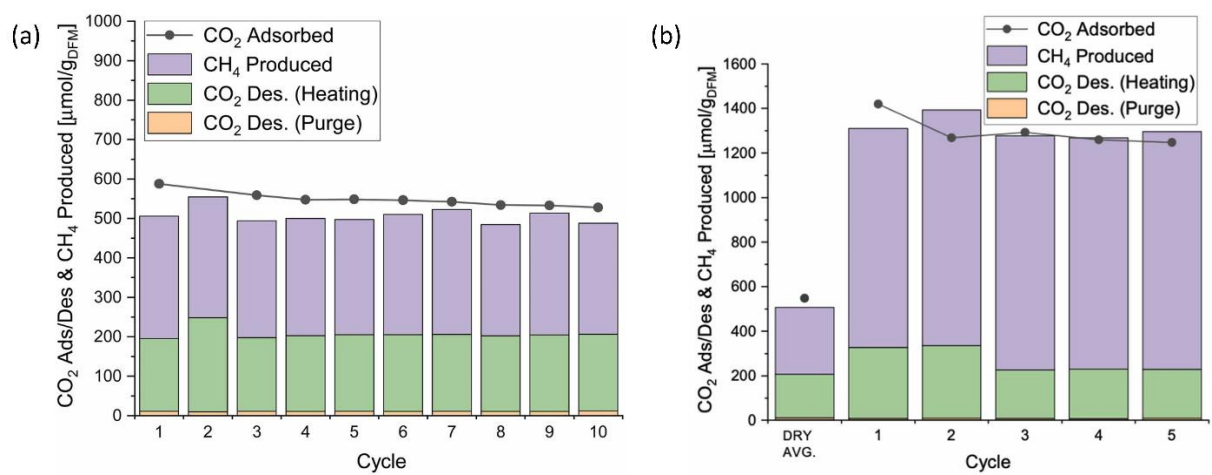

Figure 29. The results of (a) 10 cycles of adsorption, desorption and CH$_4$ production in dry air condition. (b) 5 cycles in 2 % wet air condition.

DFMs were also investigated to convert CO$_2$ to syngas by utilizing abundant base metals[292]. Researchers used a sequentially impregnated FeCrCu-K catalyst supported on hydrotalcite. The hydrotalcite support was first calcined at 600 °C to produce homogenous mixed oxides of MgO and Al$_2$O$_3$, and then the metals (Fe, Cr, and Cu) were deposited using the incipient wetness technique using nitrate precursors. A K$_2$CO$_3$ solution was impregnated over the sample after it had been dried and calcined at 500 °C, and the sample was then dried and calcined once more at the same temperature. The CO$_2$ was captured using 5.8% CO$_2$/N$_2$ and hydrogenated using pure H$_2$ at a temperature between 450 and 550 °C. Figure 30 displays the CO$_2$ and CO concentration patterns through the stages of capture and reduction. As can be observed, CO$_2$ is successfully adsorbed onto the alkali metal (K), where it forms surface carbonates. Once the reaction is switched to H$_2$, the majority of the captured CO$_2$ combines with H$_2$ over the catalytic sites (Fe, Cr, and Cu), releasing CO and some unconverted CO$_2$.

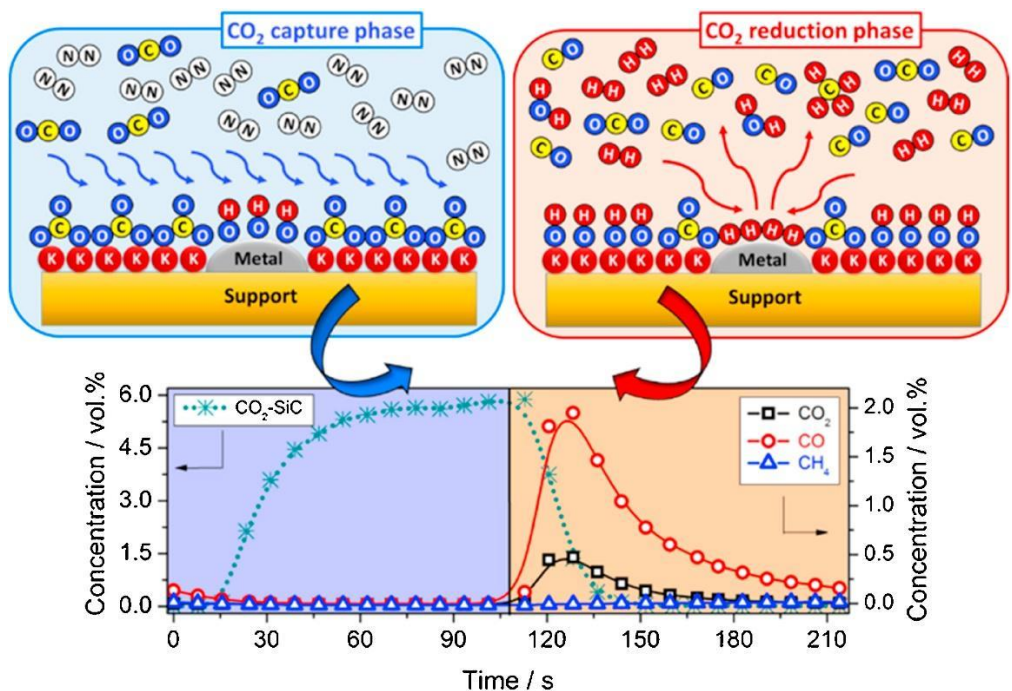

Figure 30. a plausible conversion and capture method for $CO_2$ that results in representative concentration profiles of syngas.

To comprehend the basic chemical processes and active sites involved in these, Hyakutake *et al.* [345] investigated the impact of boosting a $Cu/Al_2O_3$ catalyst with K and Ba for application in $CO_2$ collection and conversion. For the oxidative dehydrogenation of ethane (ODHE), which uses $CO_2$ as a mild oxidant, Al-Mamoori *et al.*[346] studied the use of potassium- and sodium-based calcium oxide double salts (K-Ca and Na-Ca) adsorbents physically combined with an H-ZSM-5-supported Cr catalyst. Since ethylene is one of the most essential building components in the chemical industry, this reaction is extremely alluring[347].

### 7.3 Photochemical and Photoelectrochemical $CO_2$ reduction

Due to the abundant and free availability of sunshine, photoreduction of $CO_2$—which mimics the natural process of photosynthesis—is one of the most enticing processes for $CO_2$ conversion. The fundamental mechanism leading to primary production in the biosphere is photosynthesis, which annually eliminates more than 100 billion tons of $CO_2$ from the atmosphere. There are around 7,000 billion tons of $CO_2$ in the atmosphere[348]. However, a truly biomimetic photosynthetic system, like the ones inspired by Rubisco that have been proposed for directly absorbing $CO_2$ from the air, is still a long way off from being realized[349-352].

Lewis and Nocera[353] have suggested using the photosynthetic process to transform solar energy into molecules ($H_2$, methanol, and hydrocarbons) and store them to fulfill the world's energy needs. A typical photoreduction electrode consists of a semiconducting material and photocatalysts, the majority of which are complexes of transition metals. Semiconductors absorb photons to cause excited electrons to shift from a valence band to a conducting band. These excited electrons are subsequently transported to a photocatalyst complex, which converts $CO_2$ to CO and other valuable organic molecules.

According to Kumar *et al.*[354], the photocatalytic reduction of $CO_2$ to CO on Re(bipy-But)(CO)3Cl(bipy-But = 4,4'-di-tert-butyl-2,2-bipyridine)/p-type silicon achieved a faradic efficiency of 97±3%, and a short-circuit quantum efficiency of 61% for conversion. According to calculations based on density functional theory (DFT), the nature of the binding of $CO_2$ to the anion results in the formation of a $Re(bipy-tBu)(CO)_3(CO_2)K$ complex[355], shown in Figure 31.

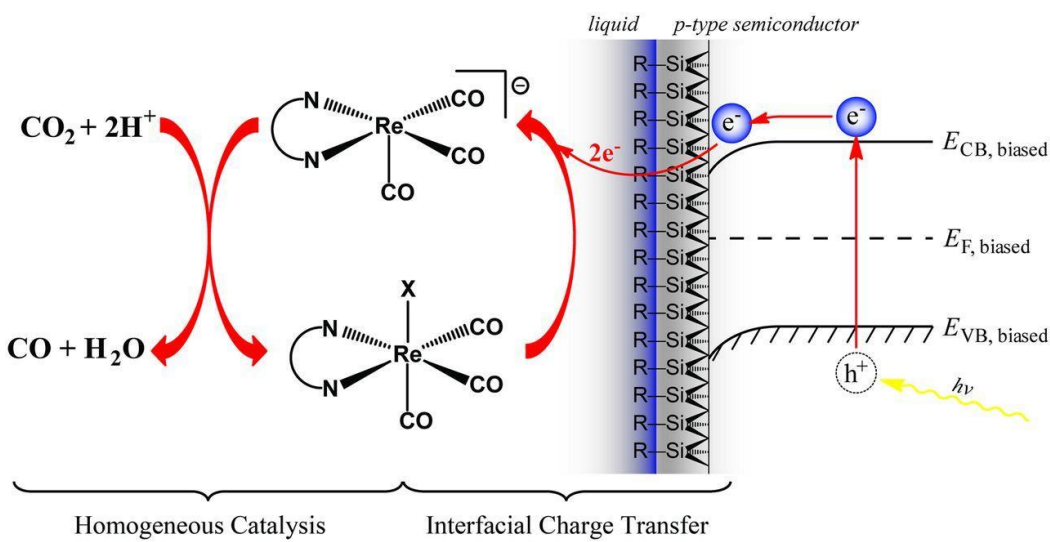

Figure 31. Schematic of a device for photoelectrochemically reducing $CO_2$ using a p-Si/ Re(bipy-tBu)(CO)₃Cl semiconductor/molecular catalyst junction.

By using a hybrid semiconductor nanowire-bacteria system, Yang's group[356-360] have developed an artificial photosynthesis that can turn $CO_2$ into goods like fuels, polymers, and drug precursors. High-surface-area silicon nanowire arrays absorb solar energy to provide reducing equivalents to the anaerobic bacterium sporomusa ovata with low overpotential (200 mV), high Faradic efficiency (up to 90%), and long-term stability (up to 200 h) for the photoelectrochemical production of acetic acid under ambient conditions (21% $O_2$)[356]. A biohybrid coculture was created by researchers from the same group for tandem and controllable $CO_2$ and $N_2$ fixation to value-added products[360].

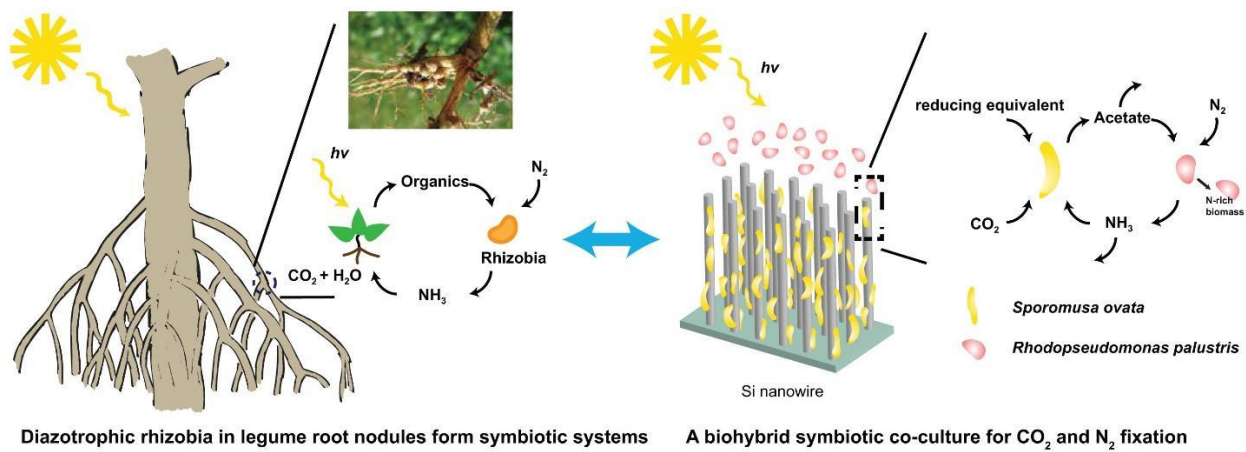

Figure 32. biohybrid coculture design using bio-inspiration. Rhizobia live in the anoxic root nodules of legumes, where the legume supplies them with organics (such as malate). The legume utilises the nitrogenous chemicals that the rhizobia fix $N_2$ to for growth. In the concept, the legume is replaced with a SiNW/S. ovata ensemble that produces acetate from $CO_2$. R. palustris utilizes the acetate as a feedstock to change $N_2$ into $NH_3$, much like rhizobia. The biohybrid platform's solar energy is represented as a ray of light, with the letter v particularly denoting the light energy. (Inset) Vigna unguiculata root nodules are shown in the image.

According to Schmid *et al.*[361], Figure 33 depicts a photosynthetic system with a solar-powered electrolyzer (e). They employed a commercially available silver-based gas diffusion cathode to solve the insufficient current densities and significantly increase the stability of the electrolyzer. With a current density of 300 $mA \cdot cm^{-2}$, it may run continuously for more than 1200 hours. In the fermentation module attached to the $CO_2$ electrolyzer, the outgoing syngas was converted with high carbon selectivity to butanol and hexanol, and this process achieved about 100% of the Faradaic efficiency. Compared to early expectations, the provided adaptive hybrid system puts artificial photosynthesis closer to industrial scale production of valuable and useful chemicals.

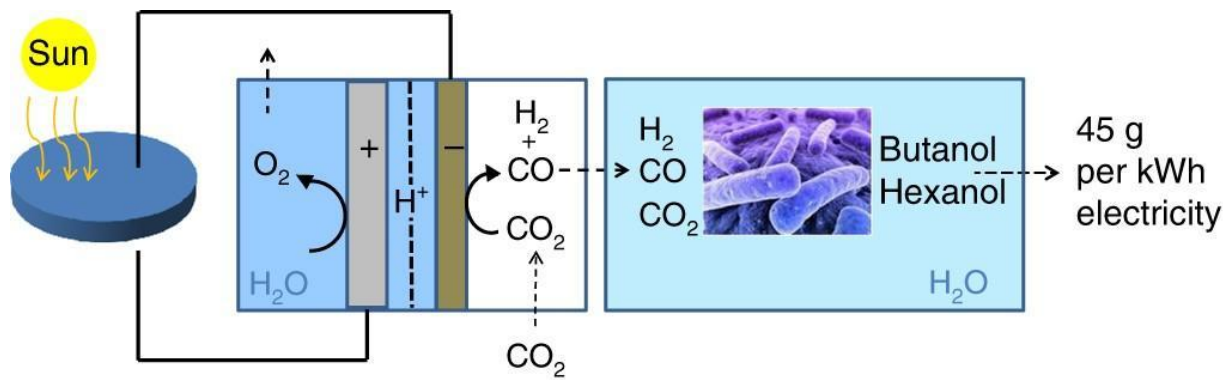

Figure 33. Sketch of the technological photosynthesis modules that produce 1-butanol and 1-hexanol from $CO_2$ and water.

## 8. Summary and Outlook:

Concerns about climate change may soon force significant reductions in $CO_2$ emissions. To address this issue, it might be required to reduce the environmental impact of fossil fuels by capturing and sequestering $CO_2$ until more affordable, eco-friendly, and abundant technologies are available. Even though there have been considerable advances in $CO_2$ capture and sequestration from the air, there are still numerous problems must be solved in the near future.

Solid sorbents are essential for removing $CO_2$ from the air. Direct Air Capture is a relatively new technology and faces numerous obstacles. A potential sorbent needs to be extremely $CO_2$ selective and easily uptake $CO_2$ from the air. It has high capacity and fast kinetics. It is essential that the procedure for releasing $CO_2$ from the sorbent is efficient and easy. The sorbent needs to endure thousands of sorbent cycles. The price must also be reasonable.

Electrified process heat is carbon-free if it is driven by zero-carbon energy generated on-site. Compared to other low-carbon process heating methods, electrification comprises a vast array of technologies that can accommodate the different uses and temperature requirements unique to the manufacturing industry. To become technically and economically practical, electrifying industrial heating processes will require overcoming a number of technological obstacles. For instance, design the procedure for electric heating to more efficiently heat the target. The optimal technology option for $CO_2$ sorbent regeneration may vary based on the operation scales and composition of the material.

Although solid amines have been extensively acknowledged as promising DAC sorbents with high $CO_2$ collection capacity and selectivity, their $CO_2$ adsorption mechanisms are still unknown. Complex computer modeling and prediction of amine reaction processes, chemical and physical characteristics, and structure-activity connections, over a spectrum of amine chemotypes, can play a significant role in complementing or enhancing limited experimental data. A synergistic combination of experimental and computational methodologies expedites the identification of an amine capture agent's ideal qualities compared to the use of experiments alone. Such an approach will expedite the development of the next generation of $CO_2$ capture materials based on amines. Computational approaches will become an increasingly valuable and complementary adjunct to experiments for the purpose of comprehending the processes of amine-$CO_2$ reactions and designing carbon capture agents with acceptable costs and toxicity.

Dual-functional materials for $CO_2$ capture and utilization are promising future prospects with the potential to create alternate routes in the sustainable chemical manufacturing and renewable energy storage industries. In addition, dual-functional materials provide additional options and the flexibility to target more valuable items that cannot be manufactured in a single step. With the exception of solid oxide electrolysis, however, the current readiness level of the electrochemical components is relatively low. Consequently, a substantial amount of effort is required to overcome crucial difficulties such as long-term stability, product selectivity, and cost-competitiveness relative to conventional industrial methods. As the development of DFMs is still in its infancy, the majority of publications have focused on the incorporation of sorbents and catalytic species via impregnation techniques; the future of the field lies in the design of well-controlled nanoarchitectures with catalytic and sorbent components in close proximity for enhanced synergy. In the field of DFMs, the generation of higher energy density products, such as alcohols and higher hydrocarbons, is a nearly unexplored route.

Overall, sorbents are essential to the success of DAC, and there is still considerable space for their enhancement. The pace of development must quicken because global warming cannot wait. Future research should assess every aspect of sorbents for DAC, including their stability, sorption kinetics, sorption capacity, selectivity, regeneration energy penalty, and cost. To alleviate the problem of global climate change, energy-efficient and low-cost $CO_2$ sorbents must be used.

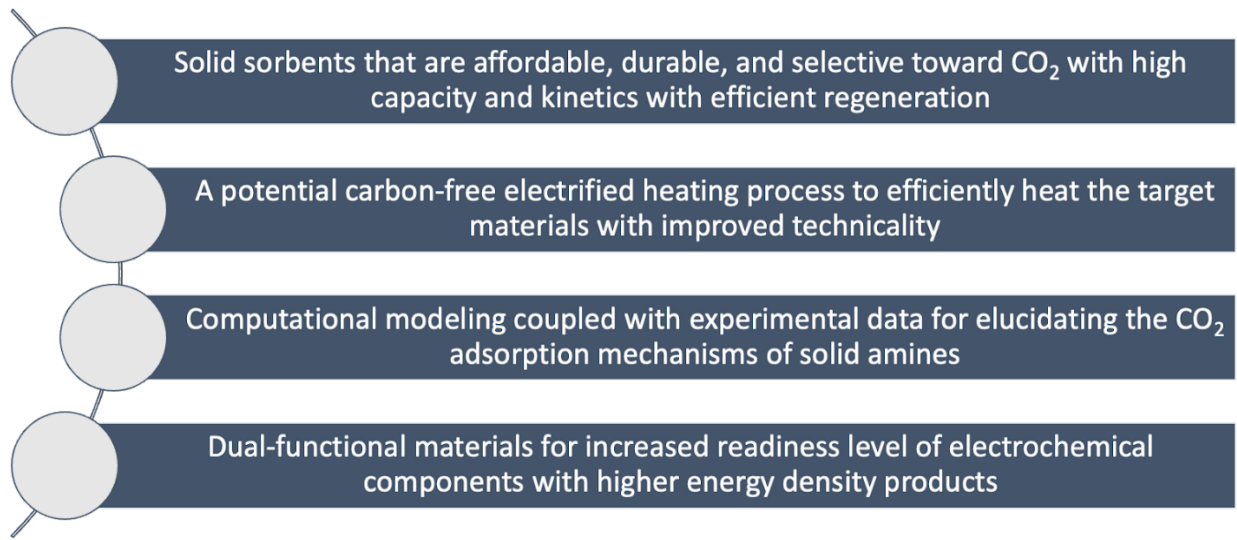

Figure 34. Outlook of the Technologies of Direct Air Capture of $CO_2$

## Acknowledgment:

Authors from Columbia University acknowledge the funding from Saudi Aramco.

## Declaration of Interests:

All authors declare no competing interests.